\documentclass[reprint,prl,aps,twocolumn,showpacs,longbibliography,floatfix,10pt,superscriptaddress]{revtex4-1}
\usepackage[dvips]{graphicx}
\usepackage{amsmath}
\usepackage{amssymb}
\usepackage[nice]{nicefrac}
\usepackage{color}

\AtBeginDocument{%
    \newwrite\bibnotes
    \def\bibnotesext{Notes.bib}
    \immediate\openout\bibnotes=\jobname\bibnotesext
    \immediate\write\bibnotes{@CONTROL{REVTEX41Control}}
    \immediate\write\bibnotes{@CONTROL{%
    apsrev41Control,author="08",editor="1",pages="1",title="0",year="1"}}
     \if@filesw
     \immediate\write\@auxout{\string\citation{apsrev41Control}}%
    \fi
}%

\begin{document}

\title{Convergence to a Gaussian by narrowing of central peak in Brownian yet non-Gaussian diffusion in disordered environments}

\author{Adrian Pacheco-Pozo}
\email{Corresponding author:\\ adrian.pacheco@physik.hu-berlin.de}
\affiliation{Institut f{\"u}r Physik, Humboldt-Universit{\"a}t zu Berlin, Newtonstra{\ss}e 15, D-12489 Berlin, Germany}
\author{Igor M. Sokolov}
\email{Corresponding author:\\ igor.sokolov@physik.hu-berlin.de}
 \affiliation{Institut f{\"u}r Physik, Humboldt-Universit{\"a}t zu Berlin, Newtonstra{\ss}e 15, D-12489 Berlin, Germany}
 \affiliation{IRIS Adlershof, Humboldt-Universit{\"a}t zu Berlin, Newtonstra{\ss}e 15, 12489 Berlin, Germany}

\date{\today}

\begin{abstract}
In usual diffusion, the concentration profile, starting from an initial distribution showing sharp features, first gets smooth and then converges to a Gaussian. By considering several examples, we show that the art of convergence to a Gaussian in diffusion in disordered media with infinite contrast may be strikingly different: sharp features of initial distribution do not smooth out at long times. This peculiarity of the strong disorder may be of importance for diagnostics of disorder in complex, e.g. biological, systems. 
\end{abstract}

\maketitle

The recent splash of interest in the precise forms of the probability density functions (PDFs) of displacements of classical particles diffusing in inhomogeneous environments was promoted by  
the experimental possibility of single particle tracking on molecular scales, see \cite{Shen}.
This lead to the discovery of an intriguing phenomenon of Brownian, yet non-Gaussian (BnG) diffusion \cite{Wang1,Wang2} (see \cite{BnG} and \cite{Burov} for more examples). 
In systems exhibiting BnG diffusion, these PDFs are strongly non-Gaussian (at least at short times, when they typically have a tent-like shape) while the mean squared displacement (MSD) grows linearly in time in the whole time domain, 
like in normal diffusion. In many systems, the PDF converges to a Gaussian at long times.
The mere fact of convergence to a Gaussian is not surprising. 
Much more interesting is \textit{how} this convergence takes place.

In several experimental situations and numerical models pertinent to BnG diffusion \cite{Wang1,Wang2,Wagner,Luo1,Post,Roichman,Pastore} 
the PDF's shape exhibits a sharp central peak which persists up to long times. 

On the other hand, the peak is absent in many pre-averaged models like continuous time random walks (CTRW) 
\cite{Barkai,Wang,Pacheco}, the diffusing diffusivity model \cite{Chubynsky} (see Fig. 1 of the work), the minimal model of BnG diffusion of Ref. \cite{BnG}, or different 
dichotomic diffusivity models \cite{Akimoto_2, Akimoto_1,Burov} (see Fig. 5 of \cite{Akimoto_1}).
Although the existence of the peak was sometimes explicitly discussed \cite{Luo1,Post}, the phenomenon did not seem to attract attention with respect to its peculiarity and possible importance.

The main statement of the present work is that the persistant peak represents a \textit{qualitative} feature of convergence to Gaussian in spacialy disordered systems: 
whereas in standard diffusion, and in CTRW and fluctuating diffusivity models the convergence to Gaussian occurs through a smoothening of the PDF, 
in disordered systems the convergence follows a distinct (and rather remarkable) pathway: the central peak does not broaden into a Gaussian, but narrows getting even sharper
under rescaling assumed by the standard definition of convergence in probability. 

Hence, the central peak is not a minor detail (``chupchik'' \cite{Haarez}), but a characteristic feature of a large class of classical disordered systems. 
We propose, therefore, that the persistence of the central peak is a distinguishing feature of systems with strong static disorder and may be used for their diagnostics.
We note that static disorder is intentionally built into the model experiments of Refs. \cite{Pastore,Roichman}, as well as in the theoretical models \cite{Luo1,Post}, but 
its presence is less evident in other experimental situations. In such cases the experimenter should search for the source of disorder, and investigate its properties. 

In what follows we show, by considering three  different models of spacially disordered systems with infinite contrast, that the phenomenon is typical for such models
independently on whether the system shows BnG diffusion in the whole time domain or only shows homogenization at longer times. In this last case, a perceptible peak persists at times 
corresponding to the homogenized behavior, when the MSD grows linearly in time. 
Such behavior essentially corresponds to a BnG diffusion
when one does not insist on the linear time dependence of the MSD \textit{in the whole time domain}, cf. \cite{Pastore}.
The overall shape of the PDF at longer times reminds of a tent standing on the top of the hill; 
the convergence to a Gaussian takes place not because the tent flattens, but because the hill grows, strikingly different 
from the predictions of the central limit theorem (CLT). 

Before turning to our models, let us discuss the predictions of CLT. 
According to the CLT, sums of independent and identically distributed (i.i.d.) random variables converge in distribution
to a Gaussian when the number of summands grows, provided the second moment of the corresponding random variables exists. 
For random variables possessing all moments, the proof is easy.
One discusses the behavior of normed sums of random variables $x_i$. One starts from centered and normalized variables $y_i = (x_i - \mu)/\sigma$ (with $\mu$ being 
the mean and $\sigma$ the variance of $x_i$), which possess zero mean and unit variance. The normed sums are then defined as 
$\xi_n = \sum_{i=1}^n y_i/\sqrt{n} =  \left[\sum_{i=1}^n (x_i - \mu)\right]/\sigma \sqrt{n}$. 
Starting from the cumulant expansion of the characteristic function for a single $y$-variable, $f(k) = \exp\left(- \frac{k^2}{2} + \sum_{m=3}^\infty \kappa_m \frac{(ik)^m}{m!} \right)$, with 
$\kappa_m$ being higher cumulants of $y_i$, one gets for a characteristic function $f_n(k)$ of $\xi_n$ the expression 
\begin{equation}
 f_n(k) = f^n\left(\frac{k}{\sqrt{n}}\right) = \exp\left(- \frac{k^2}{2} + \sum_{m=3}^\infty  \frac{(ik)^m}{m!} \frac{\kappa_m}{n^{\frac{m}{2}}} \right),
 \label{eq:ChaF}
\end{equation}
and readily infers that this converges for $n \to \infty$ to a characteristic function $\exp (- k^2/2 )$ of a Gaussian distribution with zero mean and unit variance. 

\begin{figure}[tbp]
\centering
\def\svgwidth{\columnwidth}
\begingroup%
  \makeatletter%
  \providecommand\color[2][]{%
    \errmessage{(Inkscape) Color is used for the text in Inkscape, but the package 'color.sty' is not loaded}%
    \renewcommand\color[2][]{}%
  }%
  \providecommand\transparent[1]{%
    \errmessage{(Inkscape) Transparency is used (non-zero) for the text in Inkscape, but the package 'transparent.sty' is not loaded}%
    \renewcommand\transparent[1]{}%
  }%
  \providecommand\rotatebox[2]{#2}%
  \newcommand*\fsize{\dimexpr\f@size pt\relax}%
  \newcommand*\lineheight[1]{\fontsize{\fsize}{#1\fsize}\selectfont}%
  \ifx\svgwidth\undefined%
    \setlength{\unitlength}{424.00089207bp}%
    \ifx\svgscale\undefined%
      \relax%
    \else%
      \setlength{\unitlength}{\unitlength * \real{\svgscale}}%
    \fi%
  \else%
    \setlength{\unitlength}{\svgwidth}%
  \fi%
  \global\let\svgwidth\undefined%
  \global\let\svgscale\undefined%
  \makeatother%
  \begin{picture}(1,0.76650943)%
    \lineheight{1}%
    \setlength\tabcolsep{0pt}%
    \put(0,0){\includegraphics[width=\unitlength,page=1]{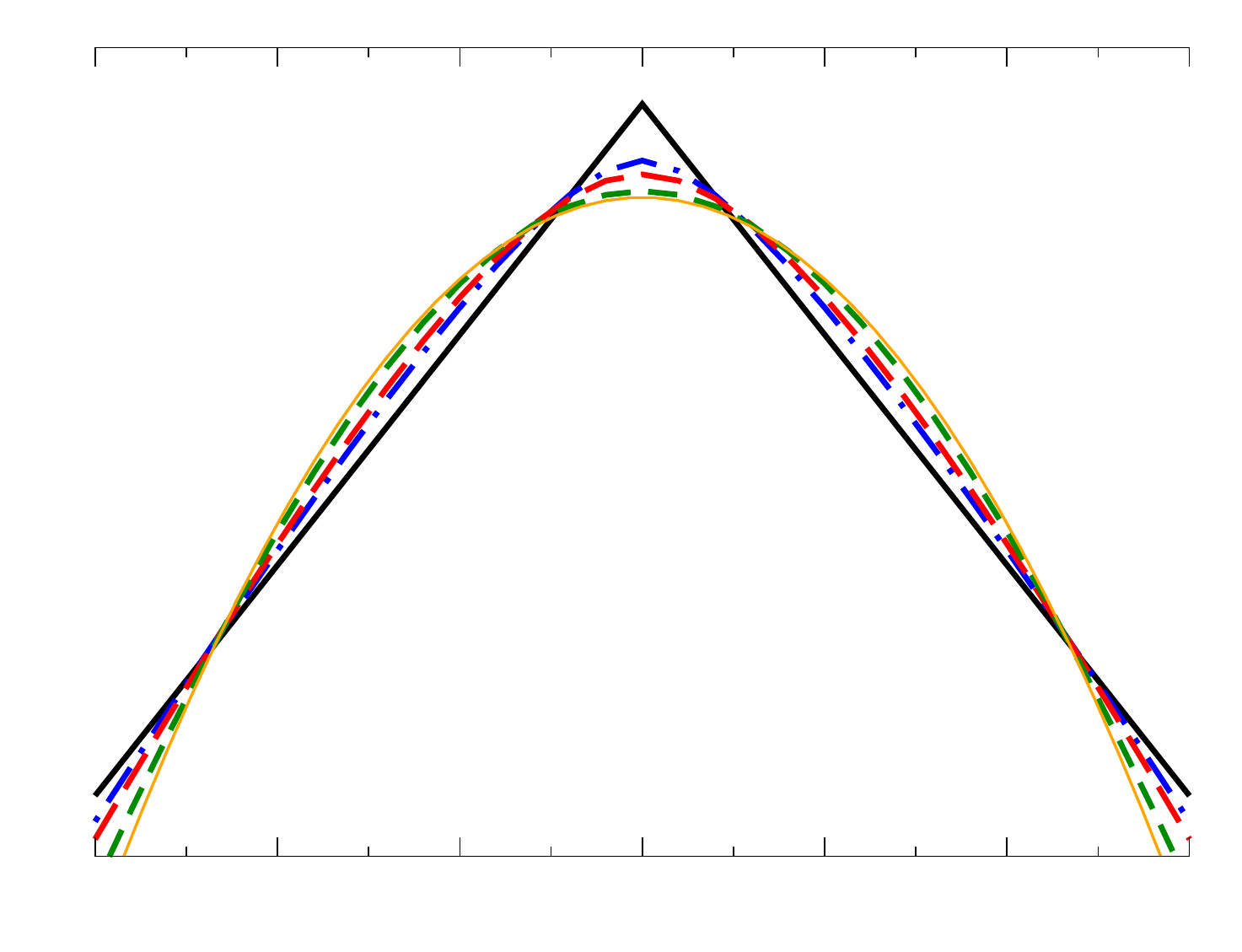}}%
    \put(0.05339623,0.04983491){\makebox(0,0)[lt]{\lineheight{1.25}\smash{\begin{tabular}[t]{l}\scriptsize $-3$\end{tabular}}}}%
    \put(0.20031061,0.04983491){\makebox(0,0)[lt]{\lineheight{1.25}\smash{\begin{tabular}[t]{l}\scriptsize $-2$\end{tabular}}}}%
    \put(0.34722476,0.04983491){\makebox(0,0)[lt]{\lineheight{1.25}\smash{\begin{tabular}[t]{l}\scriptsize $-1$\end{tabular}}}}%
    \put(0.51267689,0.04983491){\makebox(0,0)[lt]{\lineheight{1.25}\smash{\begin{tabular}[t]{l}\scriptsize $0$\end{tabular}}}}%
    \put(0.65959127,0.04983491){\makebox(0,0)[lt]{\lineheight{1.25}\smash{\begin{tabular}[t]{l}\scriptsize $1$\end{tabular}}}}%
    \put(0.80650542,0.04983491){\makebox(0,0)[lt]{\lineheight{1.25}\smash{\begin{tabular}[t]{l}\scriptsize $2$\end{tabular}}}}%
    \put(0.95341981,0.04983491){\makebox(0,0)[lt]{\lineheight{1.25}\smash{\begin{tabular}[t]{l}\scriptsize $3$\end{tabular}}}}%
    \put(0.57564009,0.02983491){\makebox(0,0)[lt]{\lineheight{1.25}\smash{\begin{tabular}[t]{l}$\xi$\end{tabular}}}}%
    \put(0,0){\includegraphics[width=\unitlength,page=2]{laplace.pdf}}%
    \put(0.00596698,0.11232972){\makebox(0,0)[lt]{\lineheight{1.25}\smash{\begin{tabular}[t]{l}\scriptsize $10^{-2}$\end{tabular}}}}%
    \put(0.00596698,0.41467901){\makebox(0,0)[lt]{\lineheight{1.25}\smash{\begin{tabular}[t]{l}\scriptsize $10^{-1}$\end{tabular}}}}%
    \put(0.02068396,0.7170283){\makebox(0,0)[lt]{\lineheight{1.25}\smash{\begin{tabular}[t]{l}\scriptsize $10^{0}$\end{tabular}}}}%
    \put(-0.01596698,0.55610849){\makebox(0,0)[lt]{\lineheight{1.25}\smash{\begin{tabular}[t]{l}$p_n(\xi)$\end{tabular}}}}%
    \put(0,0){\includegraphics[width=\unitlength,page=3]{laplace.pdf}}%
    \put(0.81479953,0.66037736){\makebox(0,0)[lt]{\lineheight{1.25}\smash{\begin{tabular}[t]{l}\tiny $n = 1$\end{tabular}}}}%
    \put(0,0){\includegraphics[width=\unitlength,page=4]{laplace.pdf}}%
    \put(0.81479953,0.63089623){\makebox(0,0)[lt]{\lineheight{1.25}\smash{\begin{tabular}[t]{l}\tiny $n = 2$\end{tabular}}}}%
    \put(0,0){\includegraphics[width=\unitlength,page=5]{laplace.pdf}}%
    \put(0.81479953,0.60141509){\makebox(0,0)[lt]{\lineheight{1.25}\smash{\begin{tabular}[t]{l}\tiny $n = 3$\end{tabular}}}}%
    \put(0,0){\includegraphics[width=\unitlength,page=6]{laplace.pdf}}%
    \put(0.81479953,0.57193396){\makebox(0,0)[lt]{\lineheight{1.25}\smash{\begin{tabular}[t]{l}\tiny $n = 10$\end{tabular}}}}%
    \put(0,0){\includegraphics[width=\unitlength,page=7]{laplace.pdf}}%
    \put(0.81479953,0.54245283){\makebox(0,0)[lt]{\lineheight{1.25}\smash{\begin{tabular}[t]{l}\tiny Gaussian\end{tabular}}}}%
    \put(0,0){\includegraphics[width=\unitlength,page=8]{laplace.pdf}}%
  \end{picture}%
\endgroup%
\caption{
The PDF of a sum of $n$ i.i.d. Laplace random variables for $n=1,2,3$, and 4, showing the ``central convergence''. Here, as also in all further figures, we use logarithmic scales.  
\label{fig:laplace}
} 
\end{figure}

It belongs to scientific folklore that the CLT is called ``central'' not only
due to its elemental importance, but also because it describes the behavior of the probability distribution in its ``center'', i.e. close to its 
mode (see e.g. \cite{LeCam}), as follows 
from the direct Edgeworth expansion, giving the corrections to the CLT
\cite{Nielsen,Hall}, see a very clear account in \cite{Petrov}. The expansion follows by direct Fourier inversion of Eq. (\ref{eq:ChaF}), so that (for continuous variables)
 \begin{equation}
 p_n(\xi) = \frac{e^{-\xi^2/2}}{\sqrt{2\pi}}\left[1 + \sum_{\nu = 1}^\infty \frac{P_{\nu}(\xi)}{n^{\nu/2}}\right].
 \label{eq:edge}
\end{equation}
Here $P_\nu(x)$ are polynomials of degree up to $m=3\nu$ with coefficients depending on the cumulants of $y_i$. The polynomial contributions are larger in the tails than in the center
of the distribution. Another property following from Eq. (\ref{eq:edge}) is smoothening: Sharp features of the PDF appear due to higher-order polynomials, whose contributions rapidly
decay with $n$. In Fig. \ref{fig:laplace} we give an example of such convergence for normed sums $\xi_n = n^{-1/2} \sum_{i=1} x_i$ of $n$ i.i.d. Laplace random variables with PDF $p(x) = (1/\sqrt{2}) \exp(-|x|/\sqrt{2})$.
The explicit forms of $p_n(\xi)$ are given in the Supplemental Material available online, Ref. \cite{Suppl}. The figure is not immediately connected with the models discussed below, 
but illustrates very well what happens under the usual mode of convergence, which we call ``central convergence'' in what follows.

This type of convergence is a reason why diffusion in homogeneous environments (say, dissolving of a droplet of ink in a quiescent homogeneous fluid) first smoothens all sharp features of initial distributions, 
and then leads to approaching Gaussian shape in the center, and then in the wings. Starting from a concentrated initial distribution, the CLT norming corresponds to 
rescaling with the root of MSD which grows as $\sqrt{t}$, so that the distribution of $\xi(t) = x(t)/\sqrt{t}$ stagnates and tends to a Gaussian when the time grows. 
This type of rescaling will be continuously used in all our examples below (Figs. 2 - 6).

\begin{figure}[tbp]
\centering
\def\svgwidth{\columnwidth}
\begingroup%
  \makeatletter%
  \providecommand\color[2][]{%
    \errmessage{(Inkscape) Color is used for the text in Inkscape, but the package 'color.sty' is not loaded}%
    \renewcommand\color[2][]{}%
  }%
  \providecommand\transparent[1]{%
    \errmessage{(Inkscape) Transparency is used (non-zero) for the text in Inkscape, but the package 'transparent.sty' is not loaded}%
    \renewcommand\transparent[1]{}%
  }%
  \providecommand\rotatebox[2]{#2}%
  \newcommand*\fsize{\dimexpr\f@size pt\relax}%
  \newcommand*\lineheight[1]{\fontsize{\fsize}{#1\fsize}\selectfont}%
  \ifx\svgwidth\undefined%
    \setlength{\unitlength}{424.00089207bp}%
    \ifx\svgscale\undefined%
      \relax%
    \else%
      \setlength{\unitlength}{\unitlength * \real{\svgscale}}%
    \fi%
  \else%
    \setlength{\unitlength}{\svgwidth}%
  \fi%
  \global\let\svgwidth\undefined%
  \global\let\svgscale\undefined%
  \makeatother%
  \begin{picture}(1,0.76650943)%
    \lineheight{1}%
    \setlength\tabcolsep{0pt}%
    \put(0,0){\includegraphics[width=\unitlength,page=1]{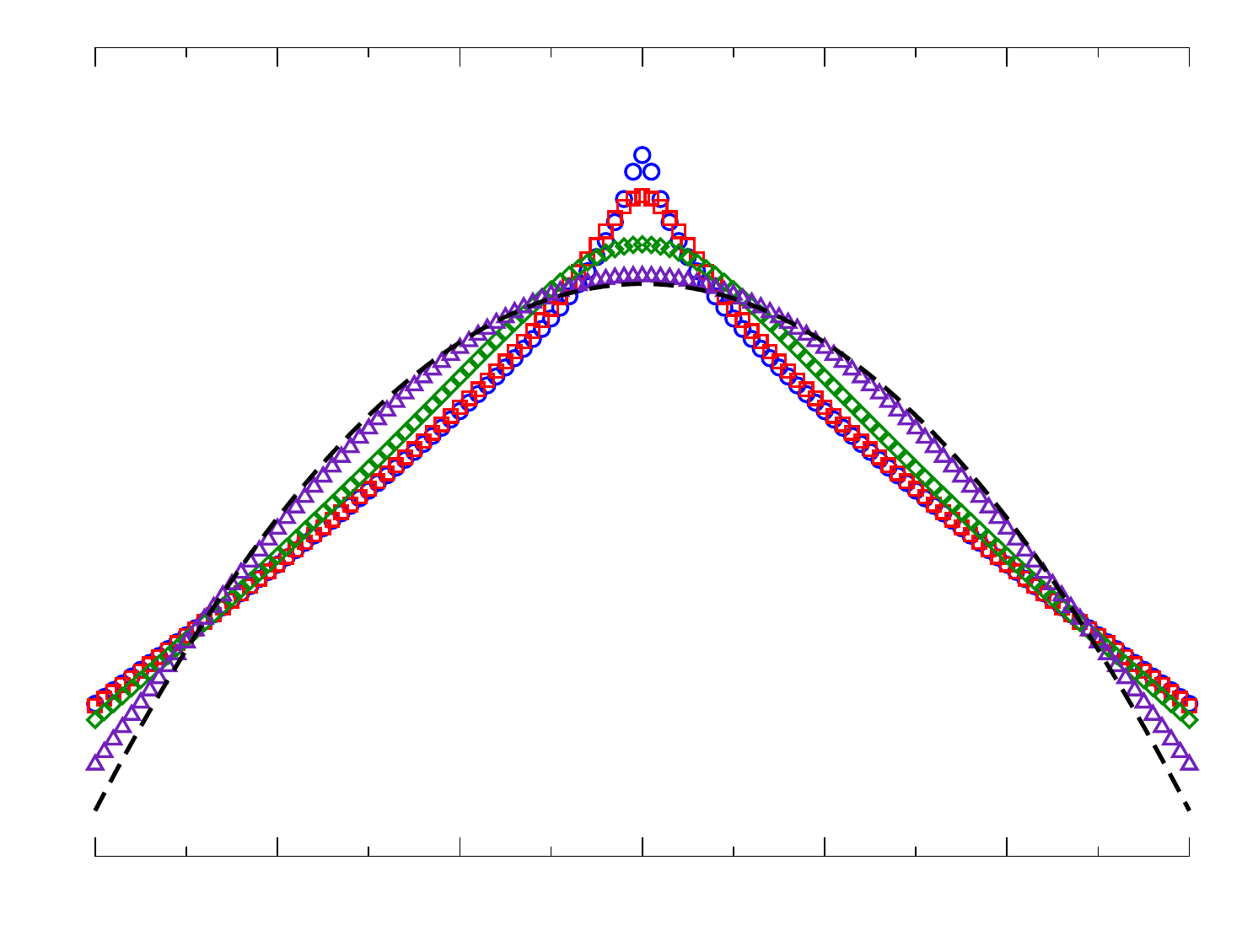}}%
    \put(0.05339623,0.04983491){\makebox(0,0)[lt]{\lineheight{1.25}\smash{\begin{tabular}[t]{l}\scriptsize $-3$\end{tabular}}}}%
    \put(0.19913137,0.04983491){\makebox(0,0)[lt]{\lineheight{1.25}\smash{\begin{tabular}[t]{l}\scriptsize $-2$\end{tabular}}}}%
    \put(0.34722476,0.04983491){\makebox(0,0)[lt]{\lineheight{1.25}\smash{\begin{tabular}[t]{l}\scriptsize $-1$\end{tabular}}}}%
    \put(0.51149764,0.04983491){\makebox(0,0)[lt]{\lineheight{1.25}\smash{\begin{tabular}[t]{l}\scriptsize $0$\end{tabular}}}}%
    \put(0.65959127,0.04983491){\makebox(0,0)[lt]{\lineheight{1.25}\smash{\begin{tabular}[t]{l}\scriptsize $1$\end{tabular}}}}%
    \put(0.80532618,0.04983491){\makebox(0,0)[lt]{\lineheight{1.25}\smash{\begin{tabular}[t]{l}\scriptsize $2$\end{tabular}}}}%
    \put(0.95341981,0.04983491){\makebox(0,0)[lt]{\lineheight{1.25}\smash{\begin{tabular}[t]{l}\scriptsize $3$\end{tabular}}}}%
    \put(0.57564009,0.02983491){\makebox(0,0)[lt]{\lineheight{1.25}\smash{\begin{tabular}[t]{l}$\xi$\end{tabular}}}}%
    \put(0,0){\includegraphics[width=\unitlength,page=2]{minimal_model.pdf}}%
    \put(-0.00360849,0.17787358){\makebox(0,0)[lt]{\lineheight{1.25}\smash{\begin{tabular}[t]{l}\scriptsize $10^{-2}$\end{tabular}}}}%
    \put(-0.00596698,0.39505142){\makebox(0,0)[lt]{\lineheight{1.25}\smash{\begin{tabular}[t]{l}\scriptsize $10^{-1}$\end{tabular}}}}%
    \put(0.01068396,0.61222901){\makebox(0,0)[lt]{\lineheight{1.25}\smash{\begin{tabular}[t]{l}\scriptsize $10^{0}$\end{tabular}}}}%
    \put(-0.01596698,0.50610849){\makebox(0,0)[lt]{\lineheight{1.25}\smash{\begin{tabular}[t]{l}$q(\xi)$\end{tabular}}}}%
    \put(0,0){\includegraphics[width=\unitlength,page=3]{minimal_model.pdf}}%
    \put(0.80078184,0.66037736){\makebox(0,0)[lt]{\lineheight{1.25}\smash{\begin{tabular}[t]{l}\tiny $t = 10^{-2}$\end{tabular}}}}%
    \put(0,0){\includegraphics[width=\unitlength,page=4]{minimal_model.pdf}}%
    \put(0.80078184,0.63089623){\makebox(0,0)[lt]{\lineheight{1.25}\smash{\begin{tabular}[t]{l}\tiny $t = 10^{-1}$\end{tabular}}}}%
    \put(0,0){\includegraphics[width=\unitlength,page=5]{minimal_model.pdf}}%
    \put(0.80078184,0.60141509){\makebox(0,0)[lt]{\lineheight{1.25}\smash{\begin{tabular}[t]{l}\tiny $t = 10^{1}$\end{tabular}}}}%
    \put(0,0){\includegraphics[width=\unitlength,page=6]{minimal_model.pdf}}%
    \put(0.80078184,0.57193396){\makebox(0,0)[lt]{\lineheight{1.25}\smash{\begin{tabular}[t]{l}\tiny $t = 10^{2}$\end{tabular}}}}%
    \put(0,0){\includegraphics[width=\unitlength,page=7]{minimal_model.pdf}}%
    \put(0.80078184,0.54245283){\makebox(0,0)[lt]{\lineheight{1.25}\smash{\begin{tabular}[t]{l}\tiny Gaussian\end{tabular}}}}%
    \put(0,0){\includegraphics[width=\unitlength,page=8]{minimal_model.pdf}}%
  \end{picture}%
\endgroup%
\caption{The PDF in the minimal model of BnG diffusion, Ref. \cite{BnG}, at different times. The initially sharp peak close to the mode smoothens and dissolves.
\label{fig:minimal}
} 
\end{figure}

\begin{figure}[tbp]
\centering
\def\svgwidth{\columnwidth}
\begingroup%
  \makeatletter%
  \providecommand\color[2][]{%
    \errmessage{(Inkscape) Color is used for the text in Inkscape, but the package 'color.sty' is not loaded}%
    \renewcommand\color[2][]{}%
  }%
  \providecommand\transparent[1]{%
    \errmessage{(Inkscape) Transparency is used (non-zero) for the text in Inkscape, but the package 'transparent.sty' is not loaded}%
    \renewcommand\transparent[1]{}%
  }%
  \providecommand\rotatebox[2]{#2}%
  \newcommand*\fsize{\dimexpr\f@size pt\relax}%
  \newcommand*\lineheight[1]{\fontsize{\fsize}{#1\fsize}\selectfont}%
  \ifx\svgwidth\undefined%
    \setlength{\unitlength}{424.00089407bp}%
    \ifx\svgscale\undefined%
      \relax%
    \else%
      \setlength{\unitlength}{\unitlength * \real{\svgscale}}%
    \fi%
  \else%
    \setlength{\unitlength}{\svgwidth}%
  \fi%
  \global\let\svgwidth\undefined%
  \global\let\svgscale\undefined%
  \makeatother%
  \begin{picture}(1,0.76650943)%
    \lineheight{1}%
    \setlength\tabcolsep{0pt}%
    \put(0,0){\includegraphics[width=\unitlength,page=1]{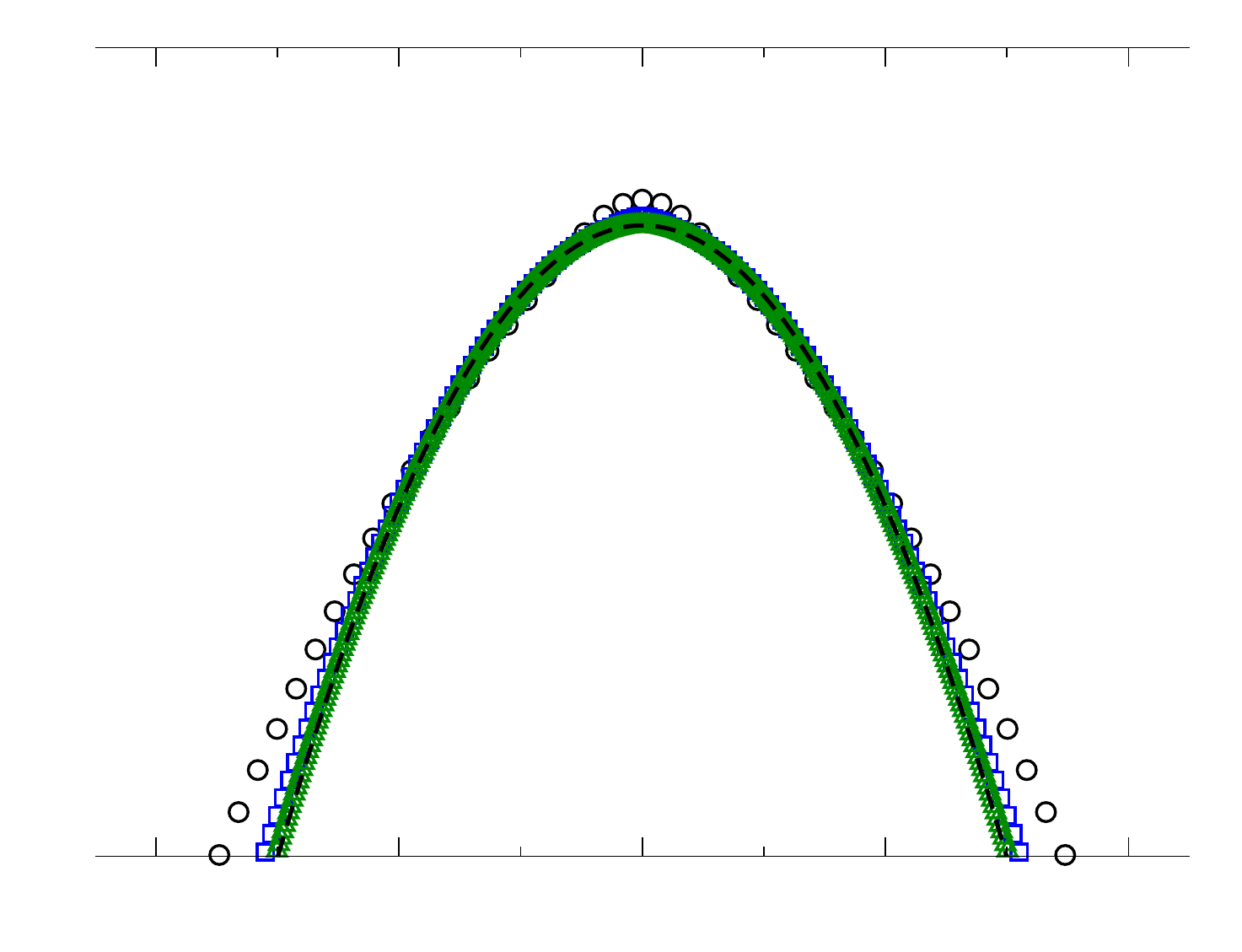}}%
    \put(0.10236769,0.04983491){\makebox(0,0)[lt]{\lineheight{1.25}\smash{\begin{tabular}[t]{l}\scriptsize $-8$\end{tabular}}}}%
    \put(0.2982533,0.04983491){\makebox(0,0)[lt]{\lineheight{1.25}\smash{\begin{tabular}[t]{l}\scriptsize $-4$\end{tabular}}}}%
    \put(0.51267689,0.04983491){\makebox(0,0)[lt]{\lineheight{1.25}\smash{\begin{tabular}[t]{l}\scriptsize $0$\end{tabular}}}}%
    \put(0.70856274,0.04983491){\makebox(0,0)[lt]{\lineheight{1.25}\smash{\begin{tabular}[t]{l}\scriptsize $4$\end{tabular}}}}%
    \put(0.90444835,0.04983491){\makebox(0,0)[lt]{\lineheight{1.25}\smash{\begin{tabular}[t]{l}\scriptsize $8$\end{tabular}}}}%
    \put(0.60564009,0.02983491){\makebox(0,0)[lt]{\lineheight{1.25}\smash{\begin{tabular}[t]{l}$\xi$\end{tabular}}}}%
    \put(0,0){\includegraphics[width=\unitlength,page=2]{ctrw.pdf}}%
    \put(0.00596698,0.06549528){\makebox(0,0)[lt]{\lineheight{1.25}\smash{\begin{tabular}[t]{l}\scriptsize $10^{-5}$\end{tabular}}}}%
    \put(0.00596698,0.19580189){\makebox(0,0)[lt]{\lineheight{1.25}\smash{\begin{tabular}[t]{l}\scriptsize $10^{-4}$\end{tabular}}}}%
    \put(0.00596698,0.32610849){\makebox(0,0)[lt]{\lineheight{1.25}\smash{\begin{tabular}[t]{l}\scriptsize $10^{-3}$\end{tabular}}}}%
    \put(0.00596698,0.45641509){\makebox(0,0)[lt]{\lineheight{1.25}\smash{\begin{tabular}[t]{l}\scriptsize $10^{-2}$\end{tabular}}}}%
    \put(0.00596698,0.5867217){\makebox(0,0)[lt]{\lineheight{1.25}\smash{\begin{tabular}[t]{l}\scriptsize $10^{-1}$\end{tabular}}}}%
    \put(0.03068396,0.7170283){\makebox(0,0)[lt]{\lineheight{1.25}\smash{\begin{tabular}[t]{l}\scriptsize $10^{0}$\end{tabular}}}}%
    \put(-0.01596698,0.38610849){\makebox(0,0)[lt]{\lineheight{1.25}\smash{\begin{tabular}[t]{l}$q(\xi)$\end{tabular}}}}%
    \put(0,0){\includegraphics[width=\unitlength,page=3]{ctrw.pdf}}%
    \put(0.75084434,0.65863797){\makebox(0,0)[lt]{\lineheight{1.25}\smash{\begin{tabular}[t]{l}\tiny $t = 10^{1}$\end{tabular}}}}%
    \put(0,0){\includegraphics[width=\unitlength,page=4]{ctrw.pdf}}%
    \put(0.75084434,0.62915684){\makebox(0,0)[lt]{\lineheight{1.25}\smash{\begin{tabular}[t]{l}\tiny $t = 10^{2}$\end{tabular}}}}%
    \put(0,0){\includegraphics[width=\unitlength,page=5]{ctrw.pdf}}%
    \put(0.75084434,0.59967571){\makebox(0,0)[lt]{\lineheight{1.25}\smash{\begin{tabular}[t]{l}\tiny $t = 10^{3}$\end{tabular}}}}%
    \put(0,0){\includegraphics[width=\unitlength,page=6]{ctrw.pdf}}%
    \put(0.39772406,0.11615566){\makebox(0,0)[lt]{\lineheight{1.25}\smash{\begin{tabular}[t]{l}\tiny $-1$\end{tabular}}}}%
    \put(0.51916274,0.11615566){\makebox(0,0)[lt]{\lineheight{1.25}\smash{\begin{tabular}[t]{l}\tiny $0$\end{tabular}}}}%
    \put(0.62942217,0.11615566){\makebox(0,0)[lt]{\lineheight{1.25}\smash{\begin{tabular}[t]{l}\tiny $1$\end{tabular}}}}%
    \put(0,0){\includegraphics[width=\unitlength,page=7]{ctrw.pdf}}%
    \put(0.31757075,0.28039316){\makebox(0,0)[lt]{\lineheight{1.25}\smash{\begin{tabular}[t]{l}\tiny $10^{-1}$\end{tabular}}}}%
    \put(0,0){\includegraphics[width=\unitlength,page=8]{ctrw.pdf}}%
  \end{picture}%
\endgroup%
\caption{The PDF in the equilibrated CTRW model being a pre-averaged approximation for the diffusivity landscape model whose (very different) behavior is represented in Fig. \ref{fig:DL}.
The inset shows the convergence close to the mode.
\label{fig:ctrw}
} 
\end{figure}

The art of convergence discussed above is also typical for many situations outside of the immediate domain of applicability of the standard CLT (subordinated models, continuous time random walks, etc.), 
given that the convergence to a Gaussian takes place at all. The two corresponding examples are given in Figs. \ref{fig:minimal} and \ref{fig:ctrw} representing the behavior of the minimal model of BnG diffusion  \cite{BnG},
and that of the equilibrated continuous time random walk (CTRW) with Pareto type II waiting time distribution possessing two lower moments,
see \cite{Suppl} for details.
These two models are close relatives of our first disordered example, the diffusivity landscape model, showing a strikingly different type of convergence. 

The integral convergence of PDFs to a limit $p_{\lim}(\xi)$ (for continuous variables and bounded PDFs) is understood in a way that for any fixed interval $[a,b]$ the integral $\int_a^b p_n(\xi) d \xi$ tends to 
$\int_a^b p_{\lim}(\xi) d \xi$ when $n$ grows (mathematically rigirous definitions are more involved and are given by the portmanteau theorem \cite{Billi}). Such convergence may mean that $p_n(\xi)$ tends to $p_{\lim}(\xi)$ everywhere within the interval (like in CLT), but may also mean that $p_n(\xi)$ stays considerably 
different from $p_{\lim}(\xi)$ on some subinterval $[a',b']$ which narrows for $n\to \infty$ (this is, as we proceed to show, the situation with the central peak), or oscillates faster and faster, 
so that the differences in the integral sense average out, as demonstrated by our last example of periodic homogenization in \cite{Suppl}.

\paragraph{The diffusion landscape model.} Now we confront the behavior corresponding to central convergence with the one obtained in the diffusivity landscape model of Ref. \cite{Post}. 
The model is chosen as our first example because it shows the phenomenon in its pure form, is easy to analyze, and also to compare it with its mean-field 
description, which is the CTRW discussed above. 
The model corresponds to a diffusion, sampled at equilibrium, in a 
discrete, correlated two-dimensional potential landscape (trap model) with the transition rates $w$ from traps distributed according to a Gamma-distribution
\begin{equation}
p(w) = 3\sqrt{\frac{3}{2 \pi}} w^{\frac{3}{2}} e^{-\frac{3}{2}w}
 \label{eq:Ito1}
\end{equation}
for the sampled diffusion coefficient set to unity, see \cite{Suppl} for details. 
This distribution of transition rates translates into the distribution of the waiting times on the sites 
\begin{equation}
 \psi(t) = \frac{45}{8}\sqrt{\frac{3}{2}} \left(t + \frac{3}{2} \right)^{-\frac{7}{2}},
 \label{eq:WTD}
\end{equation}
a Pareto type II distribution possessing two lower moments. Leaving the site, the walker goes with equal probability to each of the neighboring sites. 

\begin{figure}[tbp]
\centering
\def\svgwidth{\columnwidth}
\begingroup%
  \makeatletter%
  \providecommand\color[2][]{%
    \errmessage{(Inkscape) Color is used for the text in Inkscape, but the package 'color.sty' is not loaded}%
    \renewcommand\color[2][]{}%
  }%
  \providecommand\transparent[1]{%
    \errmessage{(Inkscape) Transparency is used (non-zero) for the text in Inkscape, but the package 'transparent.sty' is not loaded}%
    \renewcommand\transparent[1]{}%
  }%
  \providecommand\rotatebox[2]{#2}%
  \newcommand*\fsize{\dimexpr\f@size pt\relax}%
  \newcommand*\lineheight[1]{\fontsize{\fsize}{#1\fsize}\selectfont}%
  \ifx\svgwidth\undefined%
    \setlength{\unitlength}{424.00089407bp}%
    \ifx\svgscale\undefined%
      \relax%
    \else%
      \setlength{\unitlength}{\unitlength * \real{\svgscale}}%
    \fi%
  \else%
    \setlength{\unitlength}{\svgwidth}%
  \fi%
  \global\let\svgwidth\undefined%
  \global\let\svgscale\undefined%
  \makeatother%
  \begin{picture}(1,0.76650943)%
    \lineheight{1}%
    \setlength\tabcolsep{0pt}%
    \put(0,0){\includegraphics[width=\unitlength,page=1]{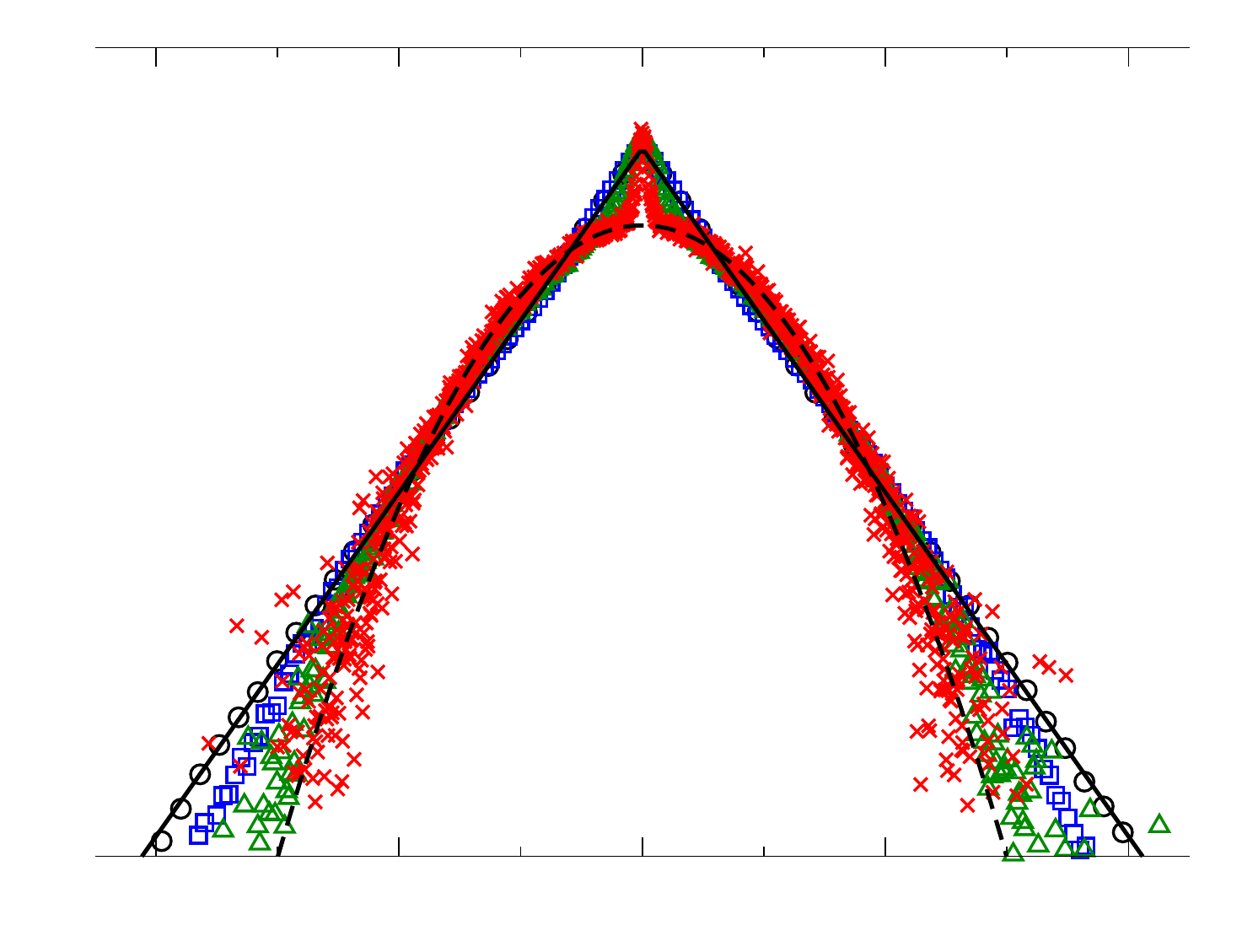}}%
    \put(0.10236769,0.04983491){\makebox(0,0)[lt]{\lineheight{1.25}\smash{\begin{tabular}[t]{l}\scriptsize $-8$\end{tabular}}}}%
    \put(0.2982533,0.04983491){\makebox(0,0)[lt]{\lineheight{1.25}\smash{\begin{tabular}[t]{l}\scriptsize $-4$\end{tabular}}}}%
    \put(0.51267689,0.04983491){\makebox(0,0)[lt]{\lineheight{1.25}\smash{\begin{tabular}[t]{l}\scriptsize $0$\end{tabular}}}}%
    \put(0.70856274,0.04983491){\makebox(0,0)[lt]{\lineheight{1.25}\smash{\begin{tabular}[t]{l}\scriptsize $4$\end{tabular}}}}%
    \put(0.90444835,0.04983491){\makebox(0,0)[lt]{\lineheight{1.25}\smash{\begin{tabular}[t]{l}\scriptsize $8$\end{tabular}}}}%
    \put(0.60564009,0.02983491){\makebox(0,0)[lt]{\lineheight{1.25}\smash{\begin{tabular}[t]{l}$\xi$\end{tabular}}}}%
    \put(0,0){\includegraphics[width=\unitlength,page=2]{DL.pdf}}%
    \put(0.00596698,0.06549528){\makebox(0,0)[lt]{\lineheight{1.25}\smash{\begin{tabular}[t]{l}\scriptsize $10^{-5}$\end{tabular}}}}%
    \put(0.00596698,0.19580189){\makebox(0,0)[lt]{\lineheight{1.25}\smash{\begin{tabular}[t]{l}\scriptsize $10^{-4}$\end{tabular}}}}%
    \put(0.00596698,0.32610849){\makebox(0,0)[lt]{\lineheight{1.25}\smash{\begin{tabular}[t]{l}\scriptsize $10^{-3}$\end{tabular}}}}%
    \put(0.00596698,0.45641509){\makebox(0,0)[lt]{\lineheight{1.25}\smash{\begin{tabular}[t]{l}\scriptsize $10^{-2}$\end{tabular}}}}%
    \put(0.00596698,0.5867217){\makebox(0,0)[lt]{\lineheight{1.25}\smash{\begin{tabular}[t]{l}\scriptsize $10^{-1}$\end{tabular}}}}%
    \put(0.03068396,0.7170283){\makebox(0,0)[lt]{\lineheight{1.25}\smash{\begin{tabular}[t]{l}\scriptsize $10^{0}$\end{tabular}}}}%
    \put(-0.01596698,0.38610849){\makebox(0,0)[lt]{\lineheight{1.25}\smash{\begin{tabular}[t]{l}$q(\xi)$\end{tabular}}}}%
    \put(0,0){\includegraphics[width=\unitlength,page=3]{DL.pdf}}%
    \put(0.75384434,0.66037736){\makebox(0,0)[lt]{\lineheight{1.25}\smash{\begin{tabular}[t]{l}\tiny $t = 10^{1}$\end{tabular}}}}%
    \put(0,0){\includegraphics[width=\unitlength,page=4]{DL.pdf}}%
    \put(0.75384434,0.63089623){\makebox(0,0)[lt]{\lineheight{1.25}\smash{\begin{tabular}[t]{l}\tiny $t = 10^{2}$\end{tabular}}}}%
    \put(0,0){\includegraphics[width=\unitlength,page=5]{DL.pdf}}%
    \put(0.75384434,0.60141509){\makebox(0,0)[lt]{\lineheight{1.25}\smash{\begin{tabular}[t]{l}\tiny $t = 10^{3}$\end{tabular}}}}%
    \put(0,0){\includegraphics[width=\unitlength,page=6]{DL.pdf}}%
    \put(0.75384434,0.57193396){\makebox(0,0)[lt]{\lineheight{1.25}\smash{\begin{tabular}[t]{l}\tiny $t = 10^{4}$\end{tabular}}}}%
    \put(0,0){\includegraphics[width=\unitlength,page=7]{DL.pdf}}%
    \put(0.39772406,0.11615566){\makebox(0,0)[lt]{\lineheight{1.25}\smash{\begin{tabular}[t]{l}\tiny $-1$\end{tabular}}}}%
    \put(0.52034198,0.11615566){\makebox(0,0)[lt]{\lineheight{1.25}\smash{\begin{tabular}[t]{l}\tiny $0$\end{tabular}}}}%
    \put(0.62942217,0.11615566){\makebox(0,0)[lt]{\lineheight{1.25}\smash{\begin{tabular}[t]{l}\tiny $1$\end{tabular}}}}%
    \put(0,0){\includegraphics[width=\unitlength,page=8]{DL.pdf}}%
    \put(0.31757075,0.20857712){\makebox(0,0)[lt]{\lineheight{1.25}\smash{\begin{tabular}[t]{l}\tiny $10^{-1}$\end{tabular}}}}%
    \put(0,0){\includegraphics[width=\unitlength,page=9]{DL.pdf}}%
  \end{picture}%
\endgroup%
\caption{
The mean PDF in the diffusivity landscape model, see \cite{Suppl} for details. The inset shows the behavior close to the mode of the distribution. The solid and dashed lines represent a Laplace and
a Gaussian PDFs, respectively. The convergence to the Gaussian takes place by narrowing, and not by lowering of the central peak like in Fig. \ref{fig:ctrw}. 
\label{fig:DL}
} 
\end{figure}

The transition rates from the traps (and thus the waiting times) are correlated with correlation length $\lambda$. We take $\lambda=10$ large enough, so that 
the model is a good approximation for a continuous situation. The details of landscape generation are given in \cite{Post, Suppl}.
At difference to the approach of Ref. \cite{Post}, here we rely on stochastic simulations of the corresponding random walk, which allows for massive parallelization and 
considerably improves statistics. The results for the scaled PDF $q(\xi)$ (see below)  of rescaled displacements $\xi(t) = x(t)/\sqrt{t}$ 
at different times are presented in Fig.~\ref{fig:DL} and show a pronounced central peak at all times. 
Shown is the mean PDF of displacements from an initial point, i.e. the one averaged over the realizations of the landscapes.
The peak narrows under rescaling (guaranteeing for the convergence to Gaussian in the integral sense), 
but does not disappear even at long times. 

\begin{figure}[tbp]
\centering
\def\svgwidth{\columnwidth}
\begingroup%
  \makeatletter%
  \providecommand\color[2][]{%
    \errmessage{(Inkscape) Color is used for the text in Inkscape, but the package 'color.sty' is not loaded}%
    \renewcommand\color[2][]{}%
  }%
  \providecommand\transparent[1]{%
    \errmessage{(Inkscape) Transparency is used (non-zero) for the text in Inkscape, but the package 'transparent.sty' is not loaded}%
    \renewcommand\transparent[1]{}%
  }%
  \providecommand\rotatebox[2]{#2}%
  \newcommand*\fsize{\dimexpr\f@size pt\relax}%
  \newcommand*\lineheight[1]{\fontsize{\fsize}{#1\fsize}\selectfont}%
  \ifx\svgwidth\undefined%
    \setlength{\unitlength}{424.00089407bp}%
    \ifx\svgscale\undefined%
      \relax%
    \else%
      \setlength{\unitlength}{\unitlength * \real{\svgscale}}%
    \fi%
  \else%
    \setlength{\unitlength}{\svgwidth}%
  \fi%
  \global\let\svgwidth\undefined%
  \global\let\svgscale\undefined%
  \makeatother%
  \begin{picture}(1,0.76650943)%
    \lineheight{1}%
    \setlength\tabcolsep{0pt}%
    \put(0,0){\includegraphics[width=\unitlength,page=1]{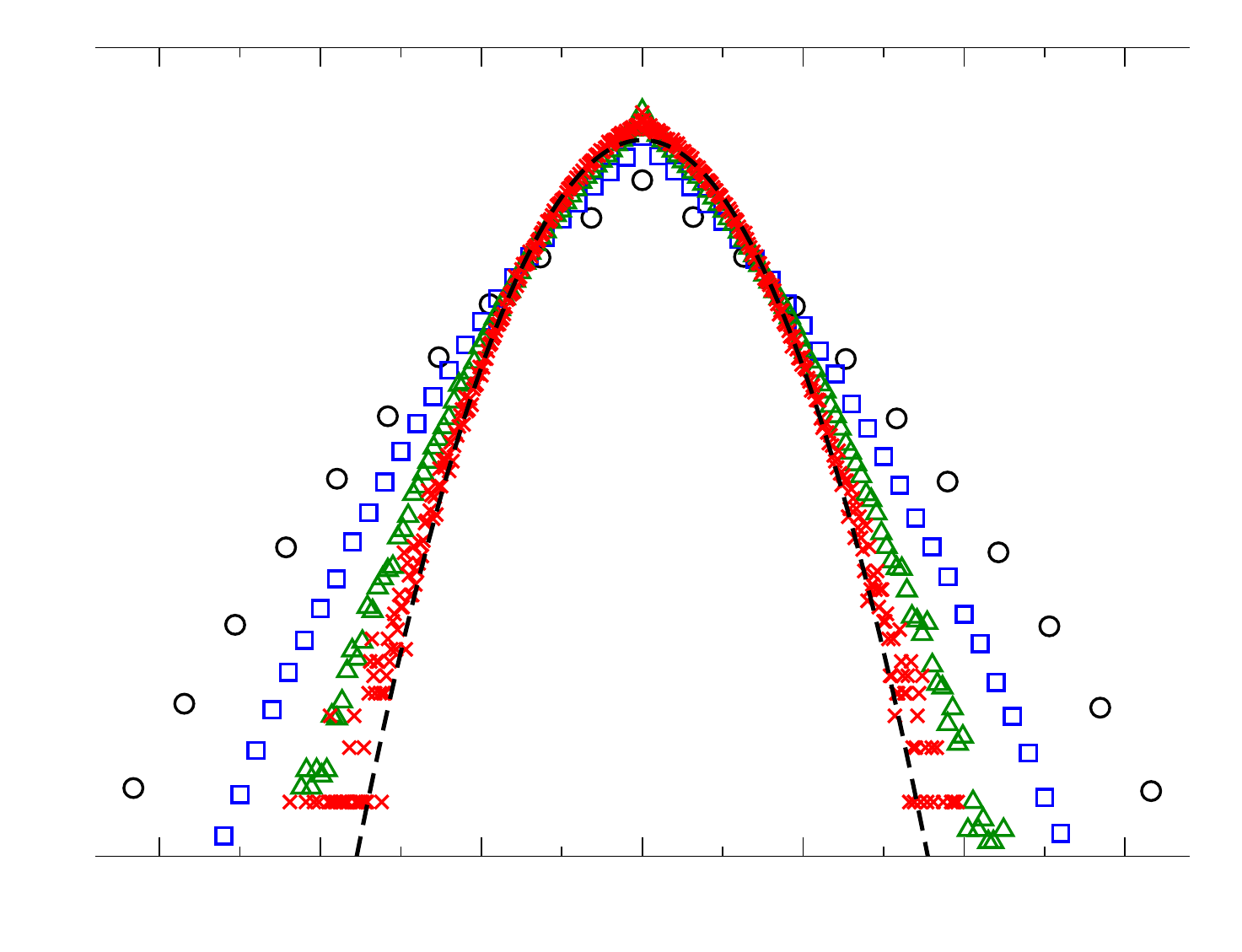}}%
    \put(0.11024835,0.04983491){\makebox(0,0)[lt]{\lineheight{1.25}\smash{\begin{tabular}[t]{l}\scriptsize $-3$\end{tabular}}}}%
    \put(0.23987854,0.04983491){\makebox(0,0)[lt]{\lineheight{1.25}\smash{\begin{tabular}[t]{l}\scriptsize $-2$\end{tabular}}}}%
    \put(0.36950896,0.04983491){\makebox(0,0)[lt]{\lineheight{1.25}\smash{\begin{tabular}[t]{l}\scriptsize $-1$\end{tabular}}}}%
    \put(0.51267689,0.04983491){\makebox(0,0)[lt]{\lineheight{1.25}\smash{\begin{tabular}[t]{l}\scriptsize $0$\end{tabular}}}}%
    \put(0.64230708,0.04983491){\makebox(0,0)[lt]{\lineheight{1.25}\smash{\begin{tabular}[t]{l}\scriptsize $1$\end{tabular}}}}%
    \put(0.7719375,0.04983491){\makebox(0,0)[lt]{\lineheight{1.25}\smash{\begin{tabular}[t]{l}\scriptsize $2$\end{tabular}}}}%
    \put(0.90156769,0.04983491){\makebox(0,0)[lt]{\lineheight{1.25}\smash{\begin{tabular}[t]{l}\scriptsize $3$\end{tabular}}}}%
    \put(0.5814009,0.02983491){\makebox(0,0)[lt]{\lineheight{1.25}\smash{\begin{tabular}[t]{l}$\xi$\end{tabular}}}}%
    \put(0,0){\includegraphics[width=\unitlength,page=2]{percolation.pdf}}%
    \put(0.00596698,0.06549528){\makebox(0,0)[lt]{\lineheight{1.25}\smash{\begin{tabular}[t]{l}\scriptsize $10^{-4}$\end{tabular}}}}%
    \put(0.00596698,0.21102028){\makebox(0,0)[lt]{\lineheight{1.25}\smash{\begin{tabular}[t]{l}\scriptsize $10^{-3}$\end{tabular}}}}%
    \put(0.00596698,0.35654528){\makebox(0,0)[lt]{\lineheight{1.25}\smash{\begin{tabular}[t]{l}\scriptsize $10^{-2}$\end{tabular}}}}%
    \put(0.00596698,0.50207028){\makebox(0,0)[lt]{\lineheight{1.25}\smash{\begin{tabular}[t]{l}\scriptsize $10^{-1}$\end{tabular}}}}%
    \put(0.02596698,0.64759528){\makebox(0,0)[lt]{\lineheight{1.25}\smash{\begin{tabular}[t]{l}\scriptsize $10^{0}$\end{tabular}}}}%
    \put(-0.01596698,0.42610849){\makebox(0,0)[lt]{\lineheight{1.25}\smash{\begin{tabular}[t]{l}$q(\xi)$\end{tabular}}}}%
    \put(0,0){\includegraphics[width=\unitlength,page=3]{percolation.pdf}}%
    \put(0.75384434,0.66037736){\makebox(0,0)[lt]{\lineheight{1.25}\smash{\begin{tabular}[t]{l}\tiny $t = 10^{1}$\end{tabular}}}}%
    \put(0,0){\includegraphics[width=\unitlength,page=4]{percolation.pdf}}%
    \put(0.75384434,0.63089623){\makebox(0,0)[lt]{\lineheight{1.25}\smash{\begin{tabular}[t]{l}\tiny $t = 10^{2}$\end{tabular}}}}%
    \put(0,0){\includegraphics[width=\unitlength,page=5]{percolation.pdf}}%
    \put(0.75384434,0.60141509){\makebox(0,0)[lt]{\lineheight{1.25}\smash{\begin{tabular}[t]{l}\tiny $t = 10^{3}$\end{tabular}}}}%
    \put(0,0){\includegraphics[width=\unitlength,page=6]{percolation.pdf}}%
    \put(0.75384434,0.57193396){\makebox(0,0)[lt]{\lineheight{1.25}\smash{\begin{tabular}[t]{l}\tiny $t = 10^{4}$\end{tabular}}}}%
    \put(0,0){\includegraphics[width=\unitlength,page=7]{percolation.pdf}}%
    \put(0.38719646,0.11615566){\makebox(0,0)[lt]{\lineheight{1.25}\smash{\begin{tabular}[t]{l}\tiny $-0.3$\end{tabular}}}}%
    \put(0.50708726,0.11615566){\makebox(0,0)[lt]{\lineheight{1.25}\smash{\begin{tabular}[t]{l}\tiny $0.0$\end{tabular}}}}%
    \put(0.61344057,0.11615566){\makebox(0,0)[lt]{\lineheight{1.25}\smash{\begin{tabular}[t]{l}\tiny $0.3$\end{tabular}}}}%
    \put(0,0){\includegraphics[width=\unitlength,page=8]{percolation.pdf}}%
    \put(0.33528774,0.23029481){\makebox(0,0)[lt]{\lineheight{1.25}\smash{\begin{tabular}[t]{l}\tiny $10^{0}$\end{tabular}}}}%
    \put(0,0){\includegraphics[width=\unitlength,page=9]{percolation.pdf}}%
  \end{picture}%
\endgroup%
\caption{The mean PDF of particles' displacements on an infinite percolation cluster.
\label{fig:per}
} 
\end{figure}

\begin{figure}[tbp]
\centering
\def\svgwidth{\columnwidth}
\begingroup%
  \makeatletter%
  \providecommand\color[2][]{%
    \errmessage{(Inkscape) Color is used for the text in Inkscape, but the package 'color.sty' is not loaded}%
    \renewcommand\color[2][]{}%
  }%
  \providecommand\transparent[1]{%
    \errmessage{(Inkscape) Transparency is used (non-zero) for the text in Inkscape, but the package 'transparent.sty' is not loaded}%
    \renewcommand\transparent[1]{}%
  }%
  \providecommand\rotatebox[2]{#2}%
  \newcommand*\fsize{\dimexpr\f@size pt\relax}%
  \newcommand*\lineheight[1]{\fontsize{\fsize}{#1\fsize}\selectfont}%
  \ifx\svgwidth\undefined%
    \setlength{\unitlength}{424.00089407bp}%
    \ifx\svgscale\undefined%
      \relax%
    \else%
      \setlength{\unitlength}{\unitlength * \real{\svgscale}}%
    \fi%
  \else%
    \setlength{\unitlength}{\svgwidth}%
  \fi%
  \global\let\svgwidth\undefined%
  \global\let\svgscale\undefined%
  \makeatother%
  \begin{picture}(1,0.76650943)%
    \lineheight{1}%
    \setlength\tabcolsep{0pt}%
    \put(0,0){\includegraphics[width=\unitlength,page=1]{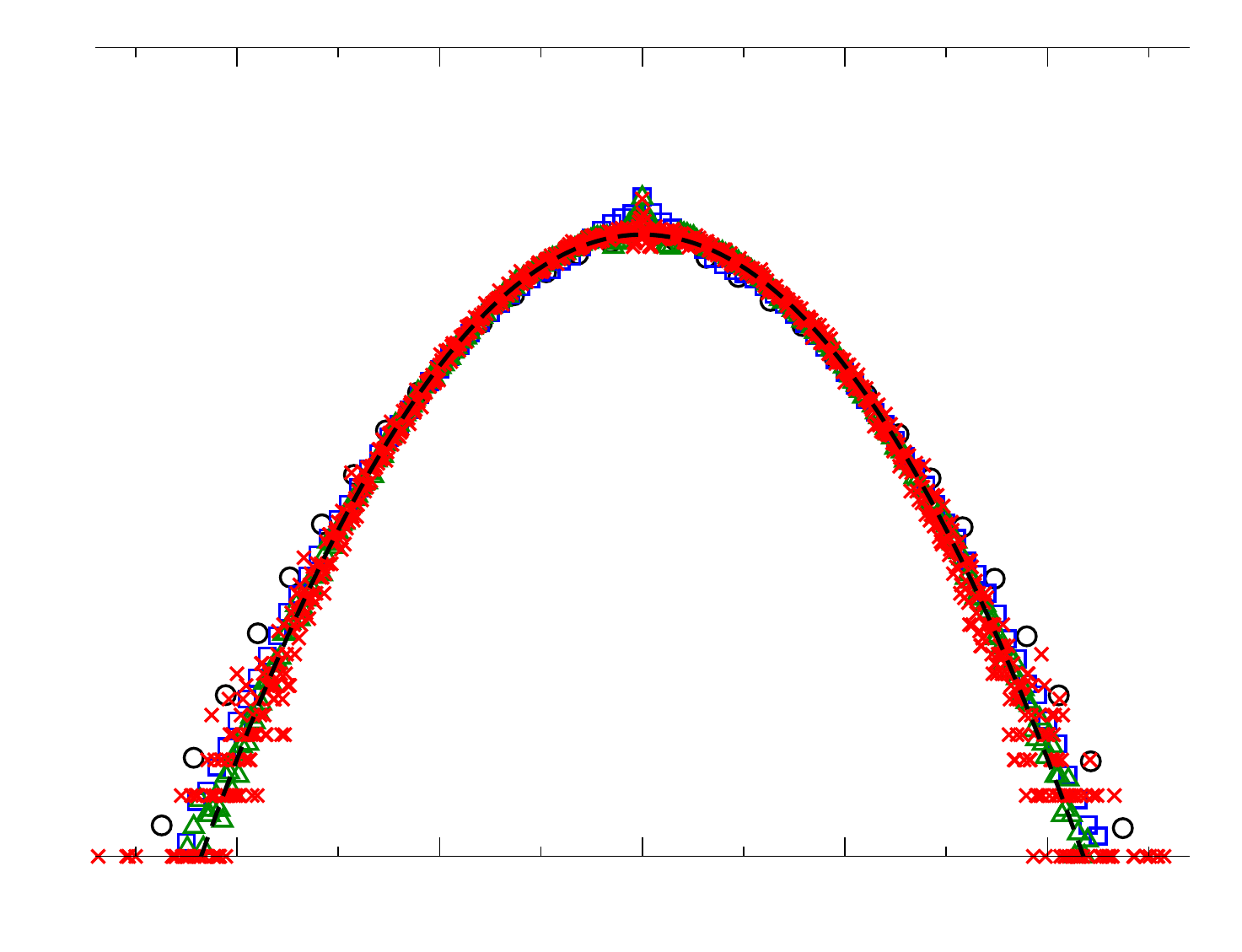}}%
    \put(0.16648373,0.04983491){\makebox(0,0)[lt]{\lineheight{1.25}\smash{\begin{tabular}[t]{l}\scriptsize $-4$\end{tabular}}}}%
    \put(0.3297217,0.04983491){\makebox(0,0)[lt]{\lineheight{1.25}\smash{\begin{tabular}[t]{l}\scriptsize $-2$\end{tabular}}}}%
    \put(0.51149764,0.04983491){\makebox(0,0)[lt]{\lineheight{1.25}\smash{\begin{tabular}[t]{l}\scriptsize $0$\end{tabular}}}}%
    \put(0.67473585,0.04983491){\makebox(0,0)[lt]{\lineheight{1.25}\smash{\begin{tabular}[t]{l}\scriptsize $2$\end{tabular}}}}%
    \put(0.83797382,0.04983491){\makebox(0,0)[lt]{\lineheight{1.25}\smash{\begin{tabular}[t]{l}\scriptsize $4$\end{tabular}}}}%
    \put(0.5914009,0.02983491){\makebox(0,0)[lt]{\lineheight{1.25}\smash{\begin{tabular}[t]{l}$\xi$\end{tabular}}}}%
    \put(0,0){\includegraphics[width=\unitlength,page=2]{spheres.pdf}}%
    \put(0.00596698,0.06431604){\makebox(0,0)[lt]{\lineheight{1.25}\smash{\begin{tabular}[t]{l}\scriptsize $10^{-4}$\end{tabular}}}}%
    \put(0.00596698,0.22719929){\makebox(0,0)[lt]{\lineheight{1.25}\smash{\begin{tabular}[t]{l}\scriptsize $10^{-3}$\end{tabular}}}}%
    \put(0.00596698,0.39008255){\makebox(0,0)[lt]{\lineheight{1.25}\smash{\begin{tabular}[t]{l}\scriptsize $10^{-2}$\end{tabular}}}}%
    \put(0.00596698,0.5529658){\makebox(0,0)[lt]{\lineheight{1.25}\smash{\begin{tabular}[t]{l}\scriptsize $10^{-1}$\end{tabular}}}}%
    \put(0.03068396,0.71584906){\makebox(0,0)[lt]{\lineheight{1.25}\smash{\begin{tabular}[t]{l}\scriptsize $10^{0}$\end{tabular}}}}%
    \put(-0.01596698,0.47610849){\makebox(0,0)[lt]{\lineheight{1.25}\smash{\begin{tabular}[t]{l}$q(\xi)$\end{tabular}}}}%
    \put(0,0){\includegraphics[width=\unitlength,page=3]{spheres.pdf}}%
    \put(0.75384434,0.66037736){\makebox(0,0)[lt]{\lineheight{1.25}\smash{\begin{tabular}[t]{l}\tiny $t = 10^{1}$\end{tabular}}}}%
    \put(0,0){\includegraphics[width=\unitlength,page=4]{spheres.pdf}}%
    \put(0.75384434,0.63089623){\makebox(0,0)[lt]{\lineheight{1.25}\smash{\begin{tabular}[t]{l}\tiny $t = 10^{2}$\end{tabular}}}}%
    \put(0,0){\includegraphics[width=\unitlength,page=5]{spheres.pdf}}%
    \put(0.75384434,0.60141509){\makebox(0,0)[lt]{\lineheight{1.25}\smash{\begin{tabular}[t]{l}\tiny $t = 10^{3}$\end{tabular}}}}%
    \put(0,0){\includegraphics[width=\unitlength,page=6]{spheres.pdf}}%
    \put(0.75384434,0.57193396){\makebox(0,0)[lt]{\lineheight{1.25}\smash{\begin{tabular}[t]{l}\tiny $t = 10^{4}$\end{tabular}}}}%
    \put(0,0){\includegraphics[width=\unitlength,page=7]{spheres.pdf}}%
    \put(0.37537925,0.11379717){\makebox(0,0)[lt]{\lineheight{1.25}\smash{\begin{tabular}[t]{l}\tiny $-0.5$\end{tabular}}}}%
    \put(0.50590802,0.11615566){\makebox(0,0)[lt]{\lineheight{1.25}\smash{\begin{tabular}[t]{l}\tiny $0.0$\end{tabular}}}}%
    \put(0.62525755,0.11379717){\makebox(0,0)[lt]{\lineheight{1.25}\smash{\begin{tabular}[t]{l}\tiny $0.5$\end{tabular}}}}%
    \put(0,0){\includegraphics[width=\unitlength,page=8]{spheres.pdf}}%
    \put(0.31257075,0.14697995){\makebox(0,0)[lt]{\lineheight{1.25}\smash{\begin{tabular}[t]{l}\tiny $10^{-1}$\end{tabular}}}}%
    \put(0,0){\includegraphics[width=\unitlength,page=9]{spheres.pdf}}%
  \end{picture}%
\endgroup%
\caption{
PDF of particles' displacement in a disordered array of solid obstacles. 
Note that the oscillations at longer times are not noise but a signature of the short-range order in a system of solid circles. 
The details of simulation and the results for the ordered counterpart of the system showing a different behavior are given in \cite{Suppl}.
\label{fig:spheres}
} 
\end{figure}

In Fig.~\ref{fig:DL}, as well as for percolation (Fig.~\ref{fig:per}) and solid obstacles problem (Fig.~\ref{fig:spheres}), 
we plot the rescaled conditional PDF $p(x) = p(x,y=0)$, as in the experimental work  \cite{Roichman}. 
The rescaled position vector in two dimensions is $(\xi,\eta)^{\mathrm{T}} =  (x,y)^{\mathrm{T}}/ \sqrt{t}$. To maintain the normalization, the PDF is rescaled according to $p(\xi,\eta) = t p(x,y)$. 
The rescaled conditional PDF is thus $q(\xi) = t p(x,y=0)$. 
The  marginal probability density $p_{\mathrm{marg}}(x) = \int p(x,y) dy$ 
shows a less pronounced (but still well visible) central peak. Plotting $rp(r)$ (as done e.g. in the early work \cite{Havlin} on percolation) fully obscures the feature. 

Since the random walk in this model is a lattice random walk with position-dependent and correlated waiting times, one can also consider a closely related equilibrated continuous time random walk (CTRW)
with waiting time density Eq.~(\ref{eq:WTD}), in which all correlations are neglected, see \cite{Suppl} for additional details. The PDF in such a CTRW, shown in Fig.~\ref{fig:ctrw},
exhibits a central convergence. The speed of convergence in the two models above, as well as in the ones below, is discussed in \cite{Suppl}.

The nature of the central peak in trap models is connected with the existence of a large number of trajectories of the random walk which never leave their
local patch. Averaging over the local diffusivities provides a sharp peak in PDF.
The estimates for the number of confined trajectories in our model follow the pattern of Ref. \cite{Luo1} and are given in \cite{Suppl}. 
The number of such confined trajectories decays as  $ \pi(t) \sim \lambda^2 / D_0 t$, and the form of the distribution in its center is tent-like, 
\begin{equation}
 p(x) \sim A \frac{\lambda}{D_0 t} - B \frac{|x|}{D_0 t} + ...
\end{equation}
(with $A$ and $B$ positive constants) forming the singular part of the overall distribution at zero. 
The presence of central peak is observed also in other classes of correlated trap models, see \cite{Luo1,Post}, also in the ones, which do now show convergence to normal diffusion, i.e. do not 
homogenize, like \cite{Luo2}.

\paragraph{Percolation model.} 

To understand, whether the ``non-central'' convergence is typical only for trap models or their close relatives, we consider a model of a very different class (barrier model, structural disorder), namely diffusion on an infinite percolation cluster well above criticality, a pet model for a strongly disordered classical system. Here, the number of confined trajectories also decays slowly, and the local properties of the system fluctuate, see \cite{Suppl} for a qualitative discussion. 

Despite an extensive search, the authors were not able to find
any simulation results for the mean PDF for such a system. The feeling is, that while concentrating on criticality, physicists  did not look carefully into the behavior in the 
homogenized regime because nothing interesting was believed to be found there.  Above criticality, the random walk on the infinite percolation cluster is known to converge for long times to
a non-degenerate, isotropic Brownian motion, see e.g. \cite{Barlow,Biskup,Mathieu}. 
Therefore, the corresponding PDF of displacements converges to a Gaussian at long times. Astonishingly, also the percolation model shows the persistence of the central peak and slow convergence 
close to the mode.

The PDF of displacements at the infinite percolation cluster well above critical concentration is shown in Fig. \ref{fig:per}.
The figure shows the PDF $q(\xi)$ for continuous-time random walks with exponential waiting times (with unit mean waiting time) on an infinite cluster of  a two-dimensional 
Bernoulli bond percolation on a simple square lattice for $p=0.55$ (critical concentration $p_c =1/2$). At small displacements, the PDF 
shows a pronounced ``chupchik'' even at long times, at which the diffusion has already homogenized (MSD grows linearly in time, see \cite{Suppl}). 
At shorter times the anomalous diffusion (subdiffusion) is observed. 
If short times are disregarded, one observes the BnG diffusion.


\paragraph{Arrangement of solid obstacles.}

This situation, especially interesting for identifying the central peak as an effect of disorder, is a close model of the experimental situation addressed in \cite{Roichman},
where the BNG diffusion was observed in an experiment on tracer diffusion in an arrangement of impenetrable obstacles (pillars) and provides the PDFs  
showing a pronounced central peak. The experiments with partly ordered systems show less pronounced peaks. 
The results of simulation of a two-dimensional variant of the model are presented in Fig. \ref{fig:spheres}. 

Like in the percolation case, the situation does not correspond to the literal BNG diffusion. Here, short- and long-time behaviors correspond to normal diffusion 
with different diffusion coefficients. At short times one encounters free diffusion, 
while the long-time diffusion coefficient in the homogenized regime is smaller, and can be estimated using known approximations \cite{Sahimi}. 
The three longer times are well in the homogenized regime.  
The overall convergence to a Gaussian is clearly seen in Fig. \ref{fig:spheres}.
The details of the simulation, and also the simulation results for a periodic arrangement of obstacles showing convergence via oscillations, are given 
in \cite{Suppl}. 

\paragraph{Discussion.} The typical pathway to convergence to a Gaussian in diffusion proceeds via smoothening sharp features of PDFs of displacements: the 
PDF, presenting at short times such sharp features first gets smooth and then, slower, approaches its final Gaussian form. In different models of 
diffusion in strongly disordered classical systems showing homogenization at large scales the art of this convergence is very different: the distribution at longer times retains a sharp 
central peak, which narrows under rescaling but does not disappear. The feature is especially pronounced
in the systems where the amount of trajectories that never leave a close neighborhood of their starting point decays slowly in time.
The feature is absent in the mean-field models, and seems to be a true sign of disorder. Finding such behavior in a physical or biological system should usher the experimentalist 
to look for the source and properties of such disorder. 

The  work  of  A.P.P.  was  financially  supported  by  “Doctoral  Programmes  in  Germany”  funded  by  the  Deutscher Akademischer   Austauschdienst   (DAAD)   (Programme   ID 57440921).

\nocite{*}


%

 \end{document}


\title{Convergence to a Gaussian by narrowing of central peak in Brownian yet non-Gaussian diffusion in disordered environments \bigskip \\ 
\underline{Supplemental material}}

\author{Adrian Pacheco-Pozo}
\affiliation{Institut f{\"u}r Physik, Humboldt-Universit{\"a}t zu Berlin, Newtonstra{\ss}e 15, D-12489 Berlin, Germany}
\author{Igor M. Sokolov}
 \affiliation{Institut f{\"u}r Physik, Humboldt-Universit{\"a}t zu Berlin, Newtonstra{\ss}e 15, D-12489 Berlin, Germany}
 \affiliation{IRIS Adlershof, Humboldt-Universit{\"a}t zu Berlin, Newtonstra{\ss}e 15, 12489 Berlin, Germany}

\date{\today}

\maketitle
 
\section{For Supplemental Material}

This supplemental material consists of the following parts: a) explicit formulae for PDFs of sums of Laplace variables, plotted in Fig. 1 of the main text;
b) explicit form of the characteristic function of the minimal model of BNG diffusion \cite{BnG_SM} whose Fourier-inversion gives the PDF for this model, plotted in Fig. 2 of the main text; 
c) some additional information on the CTRW model (Fig. 3 of the main text); d) estimates for the central peak in the diffusivity landscape model;
e) some details of numerical simulations; f) additional results of numerical simulations of percolation and solid obstacles, g) a short discussion of convergence to a Gaussian
in three disordered models and in CTRW based on the ``Gaussian parameter'',  and h) a qualitative discussion of the percolation model.

\subsection{a. Sums of Laplace variables}

An example of the central convergence is the case of summation of i.i.d. random variables having a Laplace distribution:
\begin{equation}
 p(x)=\frac{1}{\sqrt{2}} \exp(-|x|\sqrt{2}).
\end{equation}
The PDF of single variables is chosen such that second moment of this distribution is 1. 
The characteristic function of the distribution is given by
\begin{equation}
 f(k) = \int_{-\infty}^\infty \frac{1}{\sqrt{2}} \exp(-|x|\sqrt{2}) e^{ikx}dx  = \frac{1}{1+k^2/2}.
\end{equation}
The characteristic function of the distribution of the sum of $n$ such random variables is
\begin{equation}
 f_n(k) = \frac{1}{(1+k^2/2)^n},
\end{equation}
and the characteristic function of sum of renormalized variables is
\begin{equation}
 g_n(k) = \frac{1}{(1+k^2/2n)^n}.
\end{equation}
The PDF of the normed sum is
\begin{eqnarray*}
 p_n(\xi) &=& \frac{1}{2\pi} \int_{-\infty}^\infty \frac{1}{(1+k^2/2n)^n} e^{-ik\xi} dk \\
 &=&  \frac{1}{\pi} \int_0^\infty \frac{1}{(1+k^2/2n)^n} \cos(k\xi) dk.
\end{eqnarray*}
We note that the expression is an even function and consider only $\xi \geq 0$. The integral here is of the type of Eq. (2.5.6.6) of \cite{BrPr_SM}. 
To reduce it to the corresponding form we change the variable to $\kappa = k/\sqrt{2n}$, get
\begin{equation}
 p_n(\xi) = \frac{\sqrt{2n}}{\pi} \int_0^\infty \frac{1}{(1+\kappa^2)^n} \cos(\kappa \xi \sqrt{2n}) d \kappa,
\end{equation}
and denote $b= \xi \sqrt{2n}$, and $m=n-1$ to get Eq. (2.5.6.6) of \cite{BrPr_SM}. The integral reads
\begin{equation}
 \int_0^\infty \frac{\cos(\kappa b) }{(1+\kappa^2)^n} d \kappa = \frac{\pi e^{-b}}{2^{2m+1} m!} \sum_{l=0}^m \frac{(2m-k)!(2b)^k}{k!(m-k)!}.
\end{equation}
From this, the PDF of the renormalized variables reads:
\begin{eqnarray*}
 && p_n(\xi) = \\
 && \sqrt{2n} e^{-\sqrt{2n} |\xi|} \frac{1}{2^{2n-1}(n-1)!}\sum_{k=0}^{n-1} \frac{(2n-2-k)!}{k!(n-1-k)!}(2 \sqrt{2n} |\xi|)^k.
\end{eqnarray*}
The corresponding functions are plotted in Fig. 1 of the main text.

\subsection{b. The minimal model of BNG}

The characteristic function of the minimal model of Ref. \cite{BnG_SM} (with the value of parameter $n=1$) is explicitly given by the expression
\begin{eqnarray*}
 f(\mathbf{k},t) = && e^{\frac{t}{2}} \bigg[\frac{1}{2} \left(\sqrt{1+2k^2}+ \frac{1}{\sqrt{1+2k^2}} \right) \times \\
&&  \mathrm{sinh} (t\sqrt{1+2k^2})+ \mathrm{cosh}  (t\sqrt{1+2k^2})\bigg]^{-1/2}.
\end{eqnarray*}
Fig. 2 of the main text shows the distribution of $\xi$-coordinate in such a model in one dimension (which coincides with the marginal distribution
of $\xi$-coordinate in any dimension), as obtained by the numerical Fourier transform. 
In order to obtain $q(\xi) = p(\xi \sqrt{t}) \sqrt{t}$, a rescaling $k^{\prime} = k / \sqrt{t}$ should be applied, so
that $q(\xi) = \frac{1}{2\pi} \int_0^\infty \tilde{f}(k,t)e^{-ik \xi}dk$ with 
\begin{eqnarray*}
 \tilde{f}(k,t) = && e^{\frac{t}{2}} \bigg\{\frac{1}{2} \left( \frac{\sqrt{t+2k^2}} {\sqrt{t}} + \frac{\sqrt{t}}{\sqrt{t+2k^2}} \right) \times \\
&&  \mathrm{sinh} \left[\sqrt{t(t+2k^2)} \right]+ \mathrm{cosh}  \left[\sqrt{t(t+2k^2)}\right] \bigg\}^{-1/2}.
\end{eqnarray*}

\subsection{c. The CTRW model for potential landscape.}
The diffusivity landscape model corresponds to a diffusion, sampled at equilibrium, in a 
discrete, correlated two-dimensional potential landscape (trap model) with the leaving rates (inverse waiting times) $w$ on traps distributed according to a Gamma-distribution
\begin{equation}
p(w) = 3\sqrt{\frac{3}{2 \pi}} w^{\frac{3}{2}} e^{-\frac{3}{2}w}
 \label{eq:Ito2}
\end{equation}
for the sampled diffusion coefficient set to unity. The distribution of transition rates corresponds to the distribution of the waiting times $t$ on the corresponding sites.
For the given rate the conditional distribution of the waiting time is $\psi(t|w) = w \exp(-wt)$, so that
\begin{equation}
 \psi(t) = \int_0^\infty w e^{-wt} p(w) dw = \frac{45}{8}\sqrt{\frac{3}{2}} \left(t + \frac{3}{2} \right)^{-\frac{7}{2}},
 \label{eq:CTRW}
\end{equation}
which is a Pareto type II distribution possessing the finite first and the second moments, with $\langle t \rangle = 1$ and $\langle t^2 \rangle = 6$. 
The pre-averaged model for the situation, which is the CTRW, ensues when considering
the waiting times $t_i$ of all steps except for the first one as i.i.d. random variables taken to follow the PDF Eq.(\ref{eq:CTRW}). This fully neglects the two types of correlations
present in the problem: The correlations induced by the fact that the transition rates at neighboring sites are non-independent by the construction of the model, and 
the correlations along the trajectory appearing due to multiple revisiting of the same sites. The fast convergence to the Gaussian in the center of the distribution 
is clearly visible from the simulations in Fig. 3 of the main text.

To simulate the equilibrated situation \cite{KlaSok_SM}, the distribution of the first waiting time is taken to be different from all other waiting times:
\[
 \psi_1(t) = \frac{1}{\langle t \rangle} \left(1-\int_0^t \psi(t') dt' \right) = \frac{9}{4}\sqrt{\frac{3}{2}}\left(t + \frac{3}{2} \right)^{-\frac{5}{2}}.
\]
Fig. 3 of the main text shows the results of simulations with the corresponding waiting time PDFs and the two dimensional Gaussian step length distribution, which is a good approximation to the Bernoulli one 
(lattice random walk) for the scales considered.  

\subsection{d. Estimates for the central peak in the diffusivity landscape model.}

The explanation for the existence of the central peak in the diffusivity landscape model is as follows. 
Let us take a patch of the landscape with the characteristic diffusivity value $D$.
During the time $t$ the tracer typically displaces at $l \sim \sqrt{Dt}$, i.e. it leaves the patch of the characteristic dimension $\lambda$, equal to the correlation length
of the random landscape, after a time
of the order of $t \sim \lambda^2/D$. If the time $t$ is fixed, there exists a characteristic 
value of the diffusion coefficient $D_c \sim \lambda^2 t^{-1}$ such that the trajectory starting in the domain with $D<D_c$ 
does not leave this domain during the measurement time. The portion $\pi(t)$ of such trajectories among all the trajectories is of the order of 
$ \pi(t) \sim \int_0^{D_c} p(D) dD$, 
with $p(D)$ being the PDF of sampled diffusion coefficients (in our case, an exponential $p(D) = D_0^{-1} \exp(-D/D_0)$, see Ref. \cite{Post_SM}). For our case,
for $t$ long enough, we have $ \pi(t) \sim \lambda^2 / D_0 t$.

Let us concentrate only on these trajectories, which didn't yet leave their initial patch, i.e. moved in a domain with 
practically constant diffusion coefficient. Being observed at time $t$ these trajectories give the contribution to the PDF of displacements
\begin{equation}
 p(x) \sim \int_0^{D_c} \frac{1}{\sqrt{4\pi D t}} \exp \left(- \frac{x^2}{4Dt} \right) p(D) dD,
\end{equation}
which for times long enough behaves as
\begin{equation}
 p(x) \sim \frac{\pi(t)}{\lambda} \left[\frac{1}{\sqrt{\pi}} - \frac{|x|}{2 \lambda} + O(x^2) \right] \sim A \frac{\lambda}{D_0 t} - B \frac{|x|}{D_0 t} + ...
\end{equation}
forming the singular part of the overall distribution at zero. For $|x|$ small, we obtain a tent-like function with the cusp at zero. This part of the distribution is due to trajectories starting in the low-diffusivity domains
and not extending over the correlation length $\lambda$ from their origins. 
The trajectories with $x \gg \lambda$ have already adequately sampled the landscape, and contribute to the Gaussian (homogenized) wings of the distribution. 
Therefore, the whole PDF consists of a body (a ``hill'', non-singular at the origin) tending to a Gaussian, whose weight grows in the course of the time, and a tent on the hill's top, of a weight decaying as $t^{-1}$. 
Passing to the normalized variable $\xi = x/\sqrt{t}$
stabilizes the body, making it time-independent, and narrows the peak, letting it for $t \to \infty$ to converge to a single point. 

\subsection{e. Simulation details for diffusivity landscape, percolation, and solid obstacles models}

This section presents the simulation procedures employed in our work. For each of the three systems considered, the diffusivity landscape, the percolation model, and the array of hard circles, the simulation process can be split into two parts,
namely, a generation of the disordered landscapes, and a generation of the trajectories of particles moving in such disordered systems (``landscapes''). 
We first discuss the generation of the landscapes (which applies different procedures for the three cases) and then turn to modeling diffusion in these landscapes, which part of the simulation is the same for all three cases. 
All our simulations are lattice ones, with a square simulation lattice with unit lattice constant.

\subsubsection{Landscapes}

\paragraph{Diffusivity landscape.} The generation of heterogeneous diffusivity landscape follows the procedure used in Ref. \cite{Post_SM}. In the language of Ref. \cite{Post_SM}, our simulation here corresponds to an equilibrated Ito model, 
giving rise to the BnG diffusion.  The procedure consists of two steps: generating a correlated Gaussian 
random field, and its transformation to a landscape showing a single-point Gamma distribution. Hence, starting with a two-dimensional array of i.i.d. Gaussian random variables with zero mean and unit variance, one 
smoothes the landscape employing the Fourier filtering method to obtain the correlated Gaussian landscape with the correlation function chosen to be
\[
\gamma(\mathbf{r}) = \exp \left(-\frac{\mathbf{r}^2}{2 \lambda^2} \right),
\]
where $\lambda$ is the correlation length. The corresponding values of filtered Gaussian variables are then scaled to obtain a Gaussian random field $z(\mathbf{r})$ with unit variance. 

Once this correlated Gaussian random field has been generated, the diffusivity landscape is then constructed via a probability transformation, i.e. by a change of variable $z \to D$ to achieve the PDF of diffusion coefficients
to follow a Gamma-distribution
\[
p(D) = \frac{\beta^{\beta}}{\Gamma(\beta)} \frac{1}{\overline{D}} \left(\frac{D}{\overline{D}} \right)^{\beta-1} \exp \left( - \beta \frac{D}{\overline{D}} \right),
\]
where $\overline{D}$ is the mean local diffusion coefficient, $\beta$ is the shape parameter, and $\Gamma(\cdot)$ is the the Gamma function, with $\beta = (d+3)/2$ and $\overline{D} = [1 + 1/(d+1)]D_0$, with $d$ being the dimension of the space, 
and $D_0$ being the mean sampled diffusion coefficient, see \cite{Post_SM} for details. In our case $d=2$, $D_0 = 1$ we have $\beta = 5/2$ and $\overline{D} = 5 / 3$. The corresponding variable transformation is given by 
\[
 D(z) = F_\beta^{-1} \left[ \frac{1}{2} \mathrm{erfc} \left(\frac{z}{\sqrt{2}} \right) \right],
\]
with $F_\beta^{-1}$ being the inverse of the cumulative distribution function of a Gamma-distribution, $F(D) = \frac{1}{\beta} \gamma(\beta, \beta D/\overline{D}))$ with $\gamma(x,y)$ being the lower incomplete Gamma function,
which is implemented in the Fortran 90 module GammaCHI, see \cite{Gammachi_SM}. The final result of this procedure is a correlated heterogeneous diffusivity landscape. 

With the above procedure, two-dimensional square diffusivity landscapes with correlation length $\lambda = 10$ and size of $2048 \times 2048$ were generated. 
On each landscape, the diffusive motion of $10000$ particles was followed to generate a PDF. Then, a weighted mean over $10000$ realizations of the landscapes was made to obtain the mean PDF showed in Fig. 4 of the main text. 
The weights are the inverse of the diffusion coefficient at a starting site in the center of the system, where the particles begin their motion. 

\paragraph{Percolation.} For the diffusion on an infinite percolating cluster, we considered a Bernoulli bond percolation on a square lattice ($p_c = 0.5$). Finite clusters were disregarded, to avoid trapping. 
The process of generating this landscape consists of two steps: first, a large percolation system is generated by taking the bonds being intact with probability $p > p_c$, and broken otherwise. 
Once the percolation system is created, the Hoshen-Kopelman algorithm \cite{HoshKopel_SM} is used to label all clusters present in the system. 
Then the largest (``infinite'') cluster is spotted out, and the rest erased. 
The particles then start at a site of this cluster in the closest vicinity of the center of the system. The sites not belonging to the cluster are declared unavailable, and the transition probabilities to them are set to zero. 

In our simulations we use the concentration $p = 0.55$ of the intact bonds. The size of the system was $800 \times 800$. On each landscape (i.e. cluster), $10000$ particles were followed to generate the PDF in a single realization. 
Then, a mean over $5000$ realizations of the landscapes was taken to obtain the mean PDF shown in Fig. 5 of the main text.

\paragraph{Solid obstacles.} The model of the array of solid obstacles is a two-dimensional version of the experimental setup of Ref. \cite{Roichman_SM}, and corresponds to a random arrangement of the non-overlapping circles. 
The procedure to generate the landscapes is as follows. After fixing the total fraction of space $\eta$ occupied by the circles of radii $r_c$ we know the number of these circles, $N_c$, and generate the positions of their centers. 
This is done by a sequential algorithm, in which a new center is added to the lattice at random, after which one checks that its distance from all other centers is at least $R = 2r_c+2$ to ensure that the particles are not trapped. If this is not the case, the attempt is repeated. 
The procedure is repeated until the number of circles reaches $N_c$. The sites inside the circles are declared unavailable, and the corresponding system is then treated in the same way as the lattice percolation one. 

With the above process, two dimensional square landscapes were generated with unit diffusion coefficient $D_0 = 1$ in the free space, with $67600$ circles with radii $r_c = 5$ corresponding to an area fraction of circles $\eta \approx 0.3$, and a size of $4160 \times 4160$. On each landscape, $10000$ particles were used to generate the PDF. Then, a mean over $10000$ realizations of the landscapes was calculated to obtain the mean PDF showed in Fig. 6 of the main text. 

\subsubsection{Trajectories}
The diffusion of the particles is modeled as a continuous time random walk with exponentially distributed waiting time on the sites of a simulation lattice. 
The trajectories of the particles diffusing in the different landscapes were obtained using the Gillespie algorithm \cite{Toral_SM}.
Being on a site $\sigma = (i,j)$, the particle can jump to one of the four neighboring sites  $\sigma \to \delta_i$, with $\delta_1 = (i+1,j)$, $\delta_2 = (i-1,j)$, $\delta_3 = (i,j+1)$ and $\delta_4 = (i,j-1)$,
if available, with transition rates $\omega(\sigma \to \delta_i)$.
The jump rates $\omega(\sigma)$, and the probabilities to jump to different sites depend on the model. For the diffusivity landscape model, the rate of leaving the site $\sigma$ is $\omega(\sigma) = D(\sigma)$,
where $ D(\sigma)$ is the diffusion coefficient assigned to the position of the site $\sigma$, and the transition probabilities to each of the neighboring sites are $1/4$. 
For percolation and solid circle models, the transition rates from a site to each of $m$ allowed neighboring sites are taken to be the same, $\omega(\sigma \to \delta_i)= 1$, 
while the transition probabilities to forbidden sites are set to zero. This model corresponds to different jump rates at different sites, $\omega(\sigma)=1/m$, and possesses a homogeneous equilibrium.

A particle starts at $t_0=0$ from its initial position $\sigma_0$. At each simulation step $n$, corresponding to time $t_n$, the following procedure is repeated. The jump rate on the site $\sigma_n$ defines the exponentially 
distributed waiting time $\tau_n$, after which the next jump takes place. At a jump, the particle changes to one of the neighboring sites with the corresponding transition probability. The time is increased to $t_{n+1} = t_n + \tau_n$.
The procedure is repeated until the maximum simulation time $t_{\max} = 120000$ is achieved. One ends up with a list of particle's positions and respective jumping times. This list is sampled to get particle's positions 
at time intervals $\Delta t = 0.1$ which then is used for calculating the PDFs and the means. 

\subsection{f. Additional simulation results for percolation and solid obstacles} 

\begin{figure}[tbp]
\centering
\def\svgwidth{\columnwidth}
\begingroup%
  \makeatletter%
  \providecommand\color[2][]{%
    \errmessage{(Inkscape) Color is used for the text in Inkscape, but the package 'color.sty' is not loaded}%
    \renewcommand\color[2][]{}%
  }%
  \providecommand\transparent[1]{%
    \errmessage{(Inkscape) Transparency is used (non-zero) for the text in Inkscape, but the package 'transparent.sty' is not loaded}%
    \renewcommand\transparent[1]{}%
  }%
  \providecommand\rotatebox[2]{#2}%
  \newcommand*\fsize{\dimexpr\f@size pt\relax}%
  \newcommand*\lineheight[1]{\fontsize{\fsize}{#1\fsize}\selectfont}%
  \ifx\svgwidth\undefined%
    \setlength{\unitlength}{424.00089207bp}%
    \ifx\svgscale\undefined%
      \relax%
    \else%
      \setlength{\unitlength}{\unitlength * \real{\svgscale}}%
    \fi%
  \else%
    \setlength{\unitlength}{\svgwidth}%
  \fi%
  \global\let\svgwidth\undefined%
  \global\let\svgscale\undefined%
  \makeatother%
  \begin{picture}(1,0.76650943)%
    \lineheight{1}%
    \setlength\tabcolsep{0pt}%
    \put(0,0){\includegraphics[width=\unitlength,page=1]{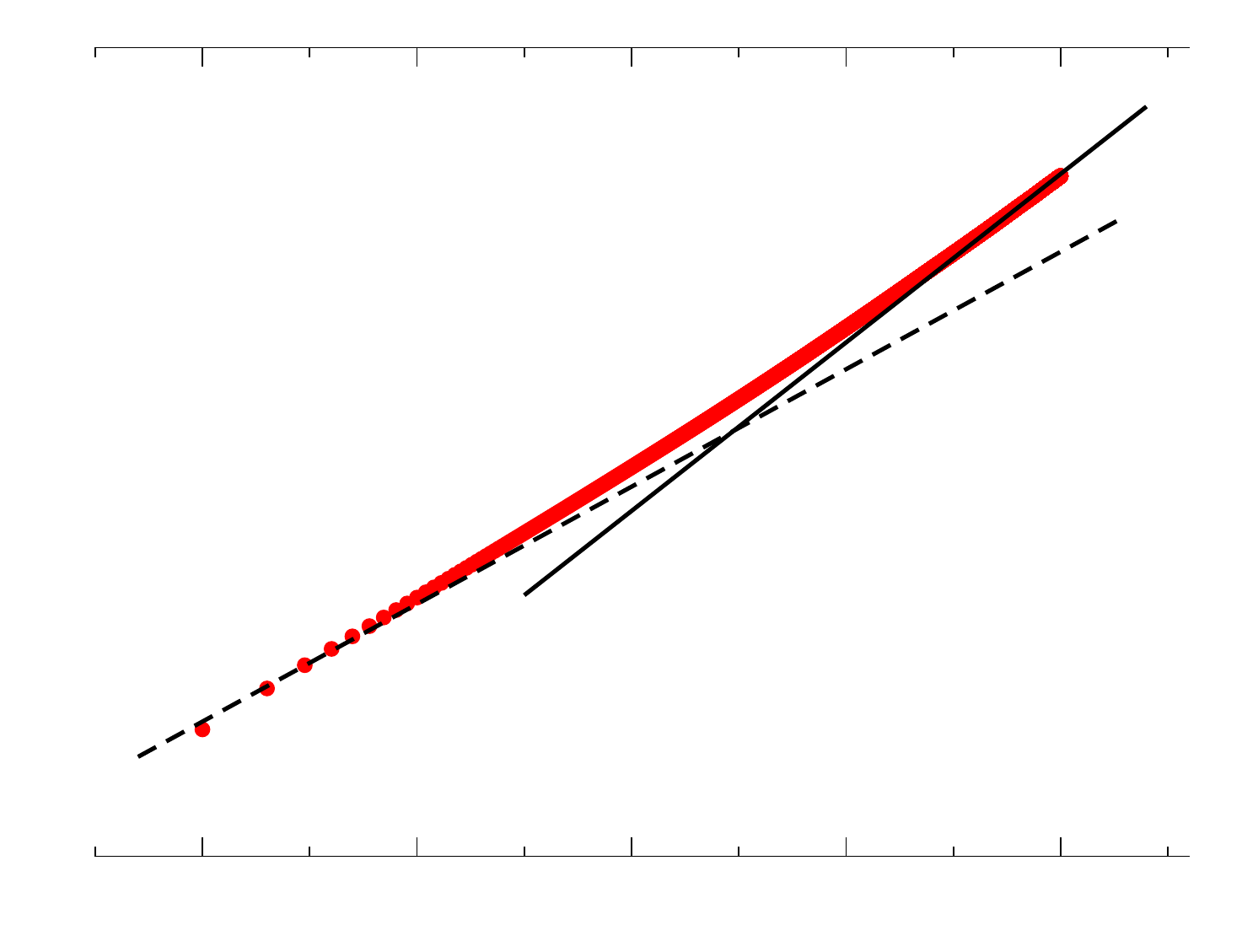}}%
    \put(0.15835425,0.04983491){\makebox(0,0)[lt]{\lineheight{1.25}\smash{\begin{tabular}[t]{l}\scriptsize $0$\end{tabular}}}}%
    \put(0.33119458,0.04983491){\makebox(0,0)[lt]{\lineheight{1.25}\smash{\begin{tabular}[t]{l}\scriptsize $1$\end{tabular}}}}%
    \put(0.50403491,0.04983491){\makebox(0,0)[lt]{\lineheight{1.25}\smash{\begin{tabular}[t]{l}\scriptsize $2$\end{tabular}}}}%
    \put(0.67687524,0.04983491){\makebox(0,0)[lt]{\lineheight{1.25}\smash{\begin{tabular}[t]{l}\scriptsize $3$\end{tabular}}}}%
    \put(0.84971557,0.04983491){\makebox(0,0)[lt]{\lineheight{1.25}\smash{\begin{tabular}[t]{l}\scriptsize $4$\end{tabular}}}}%
    \put(0.55564009,0.02983491){\makebox(0,0)[lt]{\lineheight{1.25}\smash{\begin{tabular}[t]{l}$\log_{10} t$\end{tabular}}}}%
    \put(0,0){\includegraphics[width=\unitlength,page=2]{percolation_msd.pdf}}%
    \put(0.05455189,0.1374434){\makebox(0,0)[lt]{\lineheight{1.25}\smash{\begin{tabular}[t]{l}\scriptsize $0$\end{tabular}}}}%
    \put(0.05455189,0.40891557){\makebox(0,0)[lt]{\lineheight{1.25}\smash{\begin{tabular}[t]{l}\scriptsize $2$\end{tabular}}}}%
    \put(0.05455189,0.68038774){\makebox(0,0)[lt]{\lineheight{1.25}\smash{\begin{tabular}[t]{l}\scriptsize $4$\end{tabular}}}}%
    \put(0.0495283,0.480625){\rotatebox{90}{\makebox(0,0)[lt]{\lineheight{1.25}\smash{\begin{tabular}[t]{l}$\log_{10} \langle \mathbf{r}^2 \rangle$\end{tabular}}}}}%
    \put(0,0){\includegraphics[width=\unitlength,page=3]{percolation_msd.pdf}}%
    \put(0.80545448,0.50274528){\makebox(0,0)[lt]{\lineheight{1.25}\smash{\begin{tabular}[t]{l}$\sim t^{2/d_w}$\end{tabular}}}}%
    \put(0.47796203,0.3027092){\makebox(0,0)[lt]{\lineheight{1.25}\smash{\begin{tabular}[t]{l}$\sim t$\end{tabular}}}}%
  \end{picture}%
\endgroup%
\caption{
MSD for the percolation system with $p = 0.55$ shown on double logarithmic scales. The two straight lines are drawn which correspond to different diffusion regimes. The dashed line corresponds to the subdiffusive regime,
$\langle \mathbf{r}^2 \rangle \propto t^{2/d_w}$ case with $d_w = 2.871$, and the solid line corresponds to normal diffusion.
\label{fig:msd_percolation}
} 
\end{figure}

\begin{figure}[tbp]
\centering
\def\svgwidth{\columnwidth}
\begingroup%
  \makeatletter%
  \providecommand\color[2][]{%
    \errmessage{(Inkscape) Color is used for the text in Inkscape, but the package 'color.sty' is not loaded}%
    \renewcommand\color[2][]{}%
  }%
  \providecommand\transparent[1]{%
    \errmessage{(Inkscape) Transparency is used (non-zero) for the text in Inkscape, but the package 'transparent.sty' is not loaded}%
    \renewcommand\transparent[1]{}%
  }%
  \providecommand\rotatebox[2]{#2}%
  \newcommand*\fsize{\dimexpr\f@size pt\relax}%
  \newcommand*\lineheight[1]{\fontsize{\fsize}{#1\fsize}\selectfont}%
  \ifx\svgwidth\undefined%
    \setlength{\unitlength}{424.00089407bp}%
    \ifx\svgscale\undefined%
      \relax%
    \else%
      \setlength{\unitlength}{\unitlength * \real{\svgscale}}%
    \fi%
  \else%
    \setlength{\unitlength}{\svgwidth}%
  \fi%
  \global\let\svgwidth\undefined%
  \global\let\svgscale\undefined%
  \makeatother%
  \begin{picture}(1,0.76650943)%
    \lineheight{1}%
    \setlength\tabcolsep{0pt}%
    \put(0,0){\includegraphics[width=\unitlength,page=1]{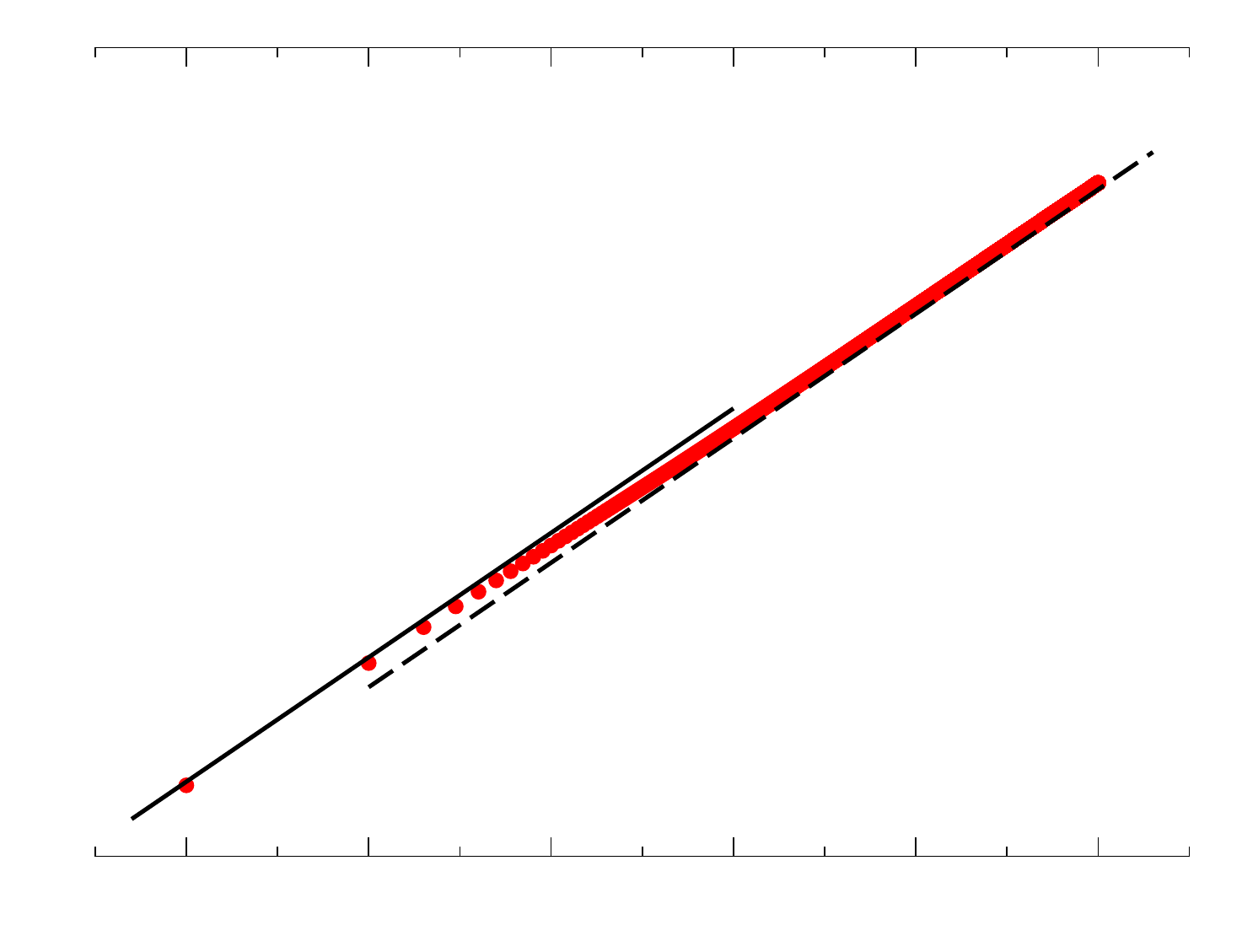}}%
    \put(0.1268533,0.04983491){\makebox(0,0)[lt]{\lineheight{1.25}\smash{\begin{tabular}[t]{l}\scriptsize $-1$\end{tabular}}}}%
    \put(0.29230542,0.04983491){\makebox(0,0)[lt]{\lineheight{1.25}\smash{\begin{tabular}[t]{l}\scriptsize $0$\end{tabular}}}}%
    \put(0.43921981,0.04983491){\makebox(0,0)[lt]{\lineheight{1.25}\smash{\begin{tabular}[t]{l}\scriptsize $1$\end{tabular}}}}%
    \put(0.58613396,0.04983491){\makebox(0,0)[lt]{\lineheight{1.25}\smash{\begin{tabular}[t]{l}\scriptsize $2$\end{tabular}}}}%
    \put(0.73304835,0.04983491){\makebox(0,0)[lt]{\lineheight{1.25}\smash{\begin{tabular}[t]{l}\scriptsize $3$\end{tabular}}}}%
    \put(0.87996274,0.04983491){\makebox(0,0)[lt]{\lineheight{1.25}\smash{\begin{tabular}[t]{l}\scriptsize $4$\end{tabular}}}}%
    \put(0.61564009,0.02983491){\makebox(0,0)[lt]{\lineheight{1.25}\smash{\begin{tabular}[t]{l}$\log_{10} t$\end{tabular}}}}%
    \put(0,0){\includegraphics[width=\unitlength,page=2]{msd_spheres.pdf}}%
    \put(0.05455189,0.16981132){\makebox(0,0)[lt]{\lineheight{1.25}\smash{\begin{tabular}[t]{l}\scriptsize $0$\end{tabular}}}}%
    \put(0.05455189,0.37028302){\makebox(0,0)[lt]{\lineheight{1.25}\smash{\begin{tabular}[t]{l}\scriptsize $2$\end{tabular}}}}%
    \put(0.05455189,0.57075472){\makebox(0,0)[lt]{\lineheight{1.25}\smash{\begin{tabular}[t]{l}\scriptsize $4$\end{tabular}}}}%
    \put(0.0495283,0.400625){\rotatebox{90}{\makebox(0,0)[lt]{\lineheight{1.25}\smash{\begin{tabular}[t]{l}$\log_{10} \langle \mathbf{r}^2 \rangle$\end{tabular}}}}}%
    \put(0,0){\includegraphics[width=\unitlength,page=3]{msd_spheres.pdf}}%
    \put(0.48432759,0.2956283){\makebox(0,0)[lt]{\lineheight{1.25}\smash{\begin{tabular}[t]{l}$4 D_{\text{EMA}} t$\end{tabular}}}}%
    \put(0.33960236,0.34873514){\makebox(0,0)[lt]{\lineheight{1.25}\smash{\begin{tabular}[t]{l}$4 D_0 t$\end{tabular}}}}%
  \end{picture}%
\endgroup%
\caption{
MSD for the array of solid obstacles with $\eta = 0.3$, on double logarithmic scales. The solid line correspond to the diffusion with the diffusion coefficient of the free diffusion, $D_0 = 1$, and the dashed line 
corresponds to diffusion with the diffusion coefficient given by the EMA, Eq. (\ref{eq:Dema}), $D_{\mathrm{EMA}} \approx 0.57$. 
\label{fig:msd_spheres}
} 
\end{figure}

Additionally to the evolution of the PDF, we also calculated the MSD for the percolation system and for the array of solid obstacles (for the diffusion landscape this was already done in Ref. \cite{Post_SM}). The MSD was obtained using the moving time average along the trajectories. The corresponding results were then averaged over the trajectories and landscapes. Fig. \ref{fig:msd_percolation} shows the MSD for the percolating cluster. 
This exhibits subdiffusion at very short times, and crosses over to normal diffusion at long ones, as discussed in the next section.  
Fig. \ref{fig:msd_spheres} shows the MSD for the array of solid obstacles. This one shows a crossover between two diffusive regimes, with a larger diffusion coefficient at short times and a lower one at long times.
This can be easily explained by the fact that particles move freely until they hit the circles; hitting the circles slows down the overall motion. The diffusion coefficient in the homogenized 
regime can be estimated using the effective medium theory (EMA), which gives
\begin{equation}
 D_{\mathrm{EMA}} = \frac{1 - 2 \eta}{ 1- \eta} D_0,
 \label{eq:Dema}
\end{equation}
where $\eta$ is the area fraction of circles in the system and $D_0$ is the free diffusion coefficient.

\begin{figure}[tbp]
\centering
\def\svgwidth{\columnwidth}
\begingroup%
  \makeatletter%
  \providecommand\color[2][]{%
    \errmessage{(Inkscape) Color is used for the text in Inkscape, but the package 'color.sty' is not loaded}%
    \renewcommand\color[2][]{}%
  }%
  \providecommand\transparent[1]{%
    \errmessage{(Inkscape) Transparency is used (non-zero) for the text in Inkscape, but the package 'transparent.sty' is not loaded}%
    \renewcommand\transparent[1]{}%
  }%
  \providecommand\rotatebox[2]{#2}%
  \newcommand*\fsize{\dimexpr\f@size pt\relax}%
  \newcommand*\lineheight[1]{\fontsize{\fsize}{#1\fsize}\selectfont}%
  \ifx\svgwidth\undefined%
    \setlength{\unitlength}{424.00089407bp}%
    \ifx\svgscale\undefined%
      \relax%
    \else%
      \setlength{\unitlength}{\unitlength * \real{\svgscale}}%
    \fi%
  \else%
    \setlength{\unitlength}{\svgwidth}%
  \fi%
  \global\let\svgwidth\undefined%
  \global\let\svgscale\undefined%
  \makeatother%
  \begin{picture}(1,0.76650943)%
    \lineheight{1}%
    \setlength\tabcolsep{0pt}%
    \put(0,0){\includegraphics[width=\unitlength,page=1]{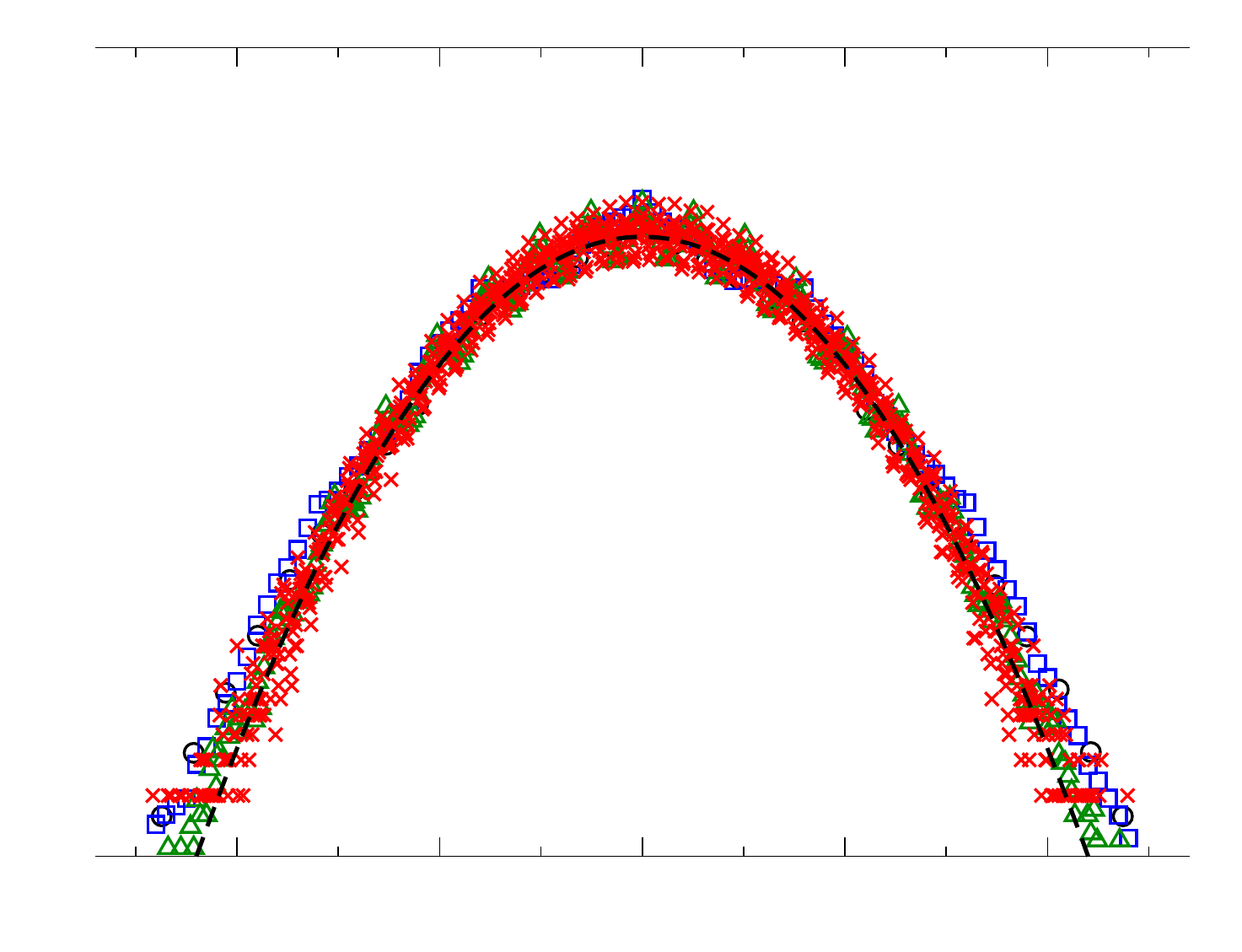}}%
    \put(0.16648373,0.04983491){\makebox(0,0)[lt]{\lineheight{1.25}\smash{\begin{tabular}[t]{l}\scriptsize $-4$\end{tabular}}}}%
    \put(0.3297217,0.04983491){\makebox(0,0)[lt]{\lineheight{1.25}\smash{\begin{tabular}[t]{l}\scriptsize $-2$\end{tabular}}}}%
    \put(0.51149764,0.04983491){\makebox(0,0)[lt]{\lineheight{1.25}\smash{\begin{tabular}[t]{l}\scriptsize $0$\end{tabular}}}}%
    \put(0.67473585,0.04983491){\makebox(0,0)[lt]{\lineheight{1.25}\smash{\begin{tabular}[t]{l}\scriptsize $2$\end{tabular}}}}%
    \put(0.83797382,0.04983491){\makebox(0,0)[lt]{\lineheight{1.25}\smash{\begin{tabular}[t]{l}\scriptsize $4$\end{tabular}}}}%
    \put(0.5914009,0.02983491){\makebox(0,0)[lt]{\lineheight{1.25}\smash{\begin{tabular}[t]{l}$\xi$\end{tabular}}}}%
    \put(0,0){\includegraphics[width=\unitlength,page=2]{spheres_order.pdf}}%
    \put(0.00596698,0.06431604){\makebox(0,0)[lt]{\lineheight{1.25}\smash{\begin{tabular}[t]{l}\scriptsize $10^{-4}$\end{tabular}}}}%
    \put(0.00596698,0.22719929){\makebox(0,0)[lt]{\lineheight{1.25}\smash{\begin{tabular}[t]{l}\scriptsize $10^{-3}$\end{tabular}}}}%
    \put(0.00596698,0.39008255){\makebox(0,0)[lt]{\lineheight{1.25}\smash{\begin{tabular}[t]{l}\scriptsize $10^{-2}$\end{tabular}}}}%
    \put(0.00596698,0.5529658){\makebox(0,0)[lt]{\lineheight{1.25}\smash{\begin{tabular}[t]{l}\scriptsize $10^{-1}$\end{tabular}}}}%
    \put(0.03068396,0.71584906){\makebox(0,0)[lt]{\lineheight{1.25}\smash{\begin{tabular}[t]{l}\scriptsize $10^{0}$\end{tabular}}}}%
    \put(-0.01596698,0.47610849){\makebox(0,0)[lt]{\lineheight{1.25}\smash{\begin{tabular}[t]{l}$q(\xi)$\end{tabular}}}}%
    \put(0,0){\includegraphics[width=\unitlength,page=3]{spheres_order.pdf}}%
    \put(0.75384434,0.66037736){\makebox(0,0)[lt]{\lineheight{1.25}\smash{\begin{tabular}[t]{l}\tiny $t = 10^{1}$\end{tabular}}}}%
    \put(0,0){\includegraphics[width=\unitlength,page=4]{spheres_order.pdf}}%
    \put(0.75384434,0.63089623){\makebox(0,0)[lt]{\lineheight{1.25}\smash{\begin{tabular}[t]{l}\tiny $t = 10^{2}$\end{tabular}}}}%
    \put(0,0){\includegraphics[width=\unitlength,page=5]{spheres_order.pdf}}%
    \put(0.75384434,0.60141509){\makebox(0,0)[lt]{\lineheight{1.25}\smash{\begin{tabular}[t]{l}\tiny $t = 10^{3}$\end{tabular}}}}%
    \put(0,0){\includegraphics[width=\unitlength,page=6]{spheres_order.pdf}}%
    \put(0.75384434,0.57193396){\makebox(0,0)[lt]{\lineheight{1.25}\smash{\begin{tabular}[t]{l}\tiny $t = 10^{4}$\end{tabular}}}}%
    \put(0,0){\includegraphics[width=\unitlength,page=7]{spheres_order.pdf}}%
    \put(0.37537925,0.11379717){\makebox(0,0)[lt]{\lineheight{1.25}\smash{\begin{tabular}[t]{l}\tiny $-0.5$\end{tabular}}}}%
    \put(0.50590802,0.11615566){\makebox(0,0)[lt]{\lineheight{1.25}\smash{\begin{tabular}[t]{l}\tiny $0.0$\end{tabular}}}}%
    \put(0.62525755,0.11379717){\makebox(0,0)[lt]{\lineheight{1.25}\smash{\begin{tabular}[t]{l}\tiny $0.5$\end{tabular}}}}%
    \put(0,0){\includegraphics[width=\unitlength,page=8]{spheres_order.pdf}}%
    \put(0.31257075,0.19009788){\makebox(0,0)[lt]{\lineheight{1.25}\smash{\begin{tabular}[t]{l}\tiny $10^{-1}$\end{tabular}}}}%
    \put(0,0){\includegraphics[width=\unitlength,page=9]{spheres_order.pdf}}%
  \end{picture}%
\endgroup%

\caption{
PDF of particles' displacement in an ordered array of solid obstacles. The PDF shows strong oscillations with the period corresponding to the unit cell of the initial landscape. 
\label{fig:order}
} 
\end{figure}

To identify the disorder as a cause of the central peak of the PDF, yet another type of landscape of solid obstacles was generated. Instead of randomly placing the centers of the circles, they are placed on an ordered lattice. 
Fig. \ref{fig:order} shows the mean PDF for such an ordered array of circles. In this case a two-dimensional square diffusivity landscape was generated, with unit free diffusion coefficient ($D_0 = 1$) and with $67600$ circles of radii $r_c = 5$ corresponding to an area fraction inside the circles $\eta \approx 0.3$. The size of the system is $4160 \times 4160$. 
The results shown in Fig. \ref{fig:order} are obtained by considering all possible inequivalent starting points within the unit cell of size $16 \times 16$ of the landscape, with a total of $10000$ trajectories.

The PDF shown in Fig.  \ref{fig:order} exhibits strong oscillations with the period corresponding to the unit cell of the initial landscape. 
In this case the convergence in distribution to a Gaussian takes place by averaging out these oscillations getting very fast under rescaling. This observation allows to 
identify the oscillations in the disordered counterpart of the system (Fig. 6 of the main text) as a consequence of the short-range order in the arrangement of non-overlapping circles:
The distance between the peaks is indeed of the order of the distance between the two peaks of the correlation function of the centers' positions.

\subsection{g. Convergence to a Gaussian}

\begin{figure}[tbp]
\centering
\def\svgwidth{\columnwidth}
\begingroup%
  \makeatletter%
  \providecommand\color[2][]{%
    \errmessage{(Inkscape) Color is used for the text in Inkscape, but the package 'color.sty' is not loaded}%
    \renewcommand\color[2][]{}%
  }%
  \providecommand\transparent[1]{%
    \errmessage{(Inkscape) Transparency is used (non-zero) for the text in Inkscape, but the package 'transparent.sty' is not loaded}%
    \renewcommand\transparent[1]{}%
  }%
  \providecommand\rotatebox[2]{#2}%
  \newcommand*\fsize{\dimexpr\f@size pt\relax}%
  \newcommand*\lineheight[1]{\fontsize{\fsize}{#1\fsize}\selectfont}%
  \ifx\svgwidth\undefined%
    \setlength{\unitlength}{424.00089407bp}%
    \ifx\svgscale\undefined%
      \relax%
    \else%
      \setlength{\unitlength}{\unitlength * \real{\svgscale}}%
    \fi%
  \else%
    \setlength{\unitlength}{\svgwidth}%
  \fi%
  \global\let\svgwidth\undefined%
  \global\let\svgscale\undefined%
  \makeatother%
  \begin{picture}(1,0.76650943)%
    \lineheight{1}%
    \setlength\tabcolsep{0pt}%
    \put(0,0){\includegraphics[width=\unitlength,page=1]{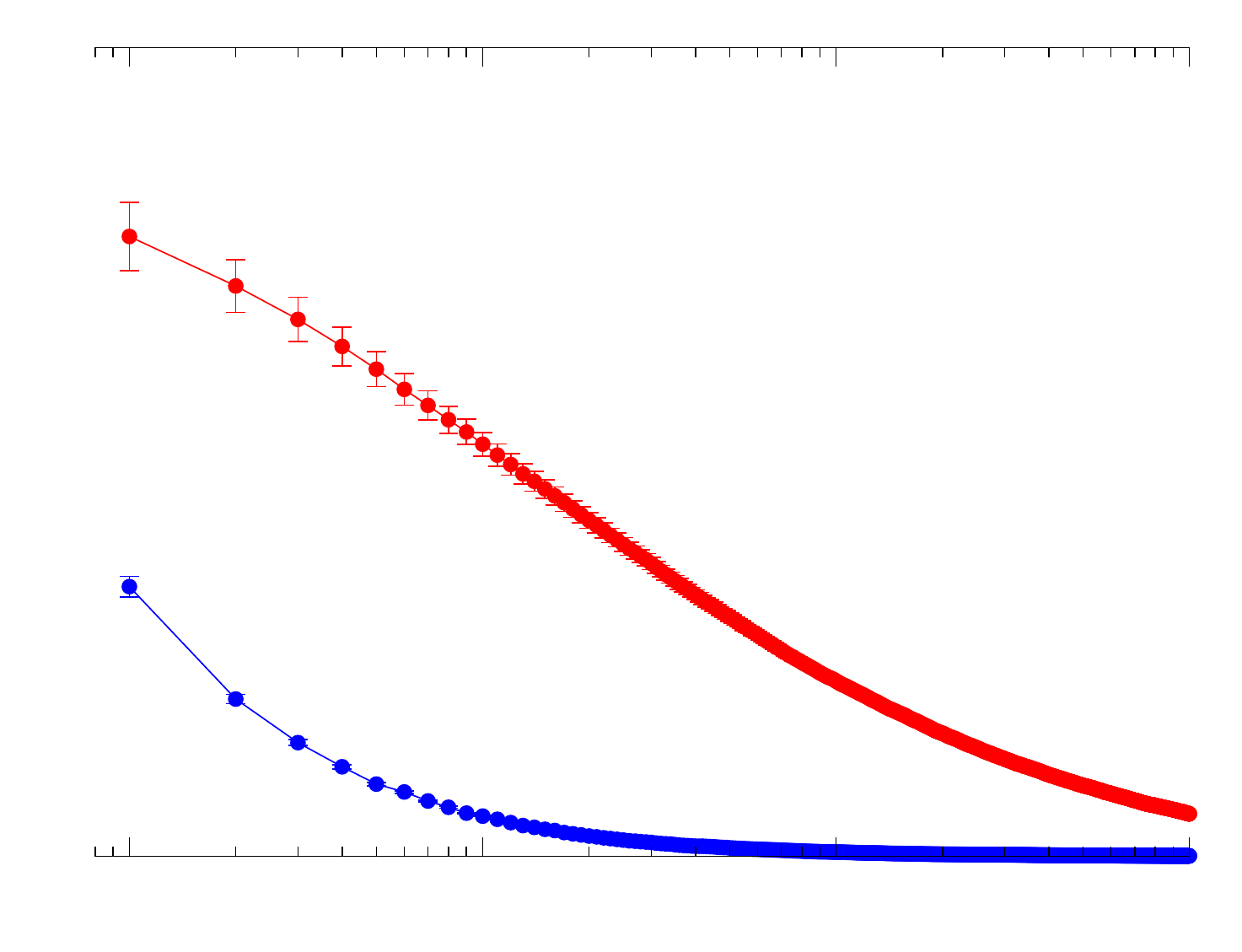}}%
    \put(0.0936217,0.03983491){\makebox(0,0)[lt]{\lineheight{1.25}\smash{\begin{tabular}[t]{l}\scriptsize $10^1$\end{tabular}}}}%
    \put(0.37353868,0.03983491){\makebox(0,0)[lt]{\lineheight{1.25}\smash{\begin{tabular}[t]{l}\scriptsize $10^2$\end{tabular}}}}%
    \put(0.65227642,0.03983491){\makebox(0,0)[lt]{\lineheight{1.25}\smash{\begin{tabular}[t]{l}\scriptsize $10^3$\end{tabular}}}}%
    \put(0.9321934,0.03983491){\makebox(0,0)[lt]{\lineheight{1.25}\smash{\begin{tabular}[t]{l}\scriptsize $10^4$\end{tabular}}}}%
    \put(0.53353868,0.01983491){\makebox(0,0)[lt]{\lineheight{1.25}\smash{\begin{tabular}[t]{l}$t$\end{tabular}}}}%
    \put(0,0){\includegraphics[width=\unitlength,page=2]{Kurtosis_1.pdf}}%
    \put(0.02804245,0.06957547){\makebox(0,0)[lt]{\lineheight{1.25}\smash{\begin{tabular}[t]{l}\scriptsize $0.0$\end{tabular}}}}%
    \put(0.02804245,0.28675307){\makebox(0,0)[lt]{\lineheight{1.25}\smash{\begin{tabular}[t]{l}\scriptsize $0.5$\end{tabular}}}}%
    \put(0.02804245,0.5039309){\makebox(0,0)[lt]{\lineheight{1.25}\smash{\begin{tabular}[t]{l}\scriptsize $1.0$\end{tabular}}}}%
    \put(0.02804245,0.72110849){\makebox(0,0)[lt]{\lineheight{1.25}\smash{\begin{tabular}[t]{l}\scriptsize $1.5$\end{tabular}}}}%
    \put(-0.01301887,0.390625){\rotatebox{0}{\makebox(0,0)[lt]{\lineheight{1.25}\smash{\begin{tabular}[t]{l}$\alpha_2(t)$\end{tabular}}}}}%
    \put(0,0){\includegraphics[width=\unitlength,page=3]{Kurtosis_1.pdf}}%
    \put(0.73325189,0.66037736){\makebox(0,0)[lt]{\lineheight{1.25}\smash{\begin{tabular}[t]{l}\tiny Diff. Land.\end{tabular}}}}%
    \put(0,0){\includegraphics[width=\unitlength,page=4]{Kurtosis_1.pdf}}%
    \put(0.73325189,0.63089623){\makebox(0,0)[lt]{\lineheight{1.25}\smash{\begin{tabular}[t]{l}\tiny CTRW\end{tabular}}}}%
    \put(0,0){\includegraphics[width=\unitlength,page=5]{Kurtosis_1.pdf}}%
  \end{picture}%
\endgroup%
\caption{Time dependence of the reduced kurtosis $\alpha_2(t)$ for the diffusivity landscape model and for the correspondng CTRW model. 
The deviations from Gaussian are larger and the convergence to $\alpha_2 =0$ is slower for the motion in the landscape than in the CTRW. 
\label{fig:Kurtosis_1}
}  
\end{figure}

\begin{figure}[tbp]
\centering
\def\svgwidth{\columnwidth}
\begingroup%
  \makeatletter%
  \providecommand\color[2][]{%
    \errmessage{(Inkscape) Color is used for the text in Inkscape, but the package 'color.sty' is not loaded}%
    \renewcommand\color[2][]{}%
  }%
  \providecommand\transparent[1]{%
    \errmessage{(Inkscape) Transparency is used (non-zero) for the text in Inkscape, but the package 'transparent.sty' is not loaded}%
    \renewcommand\transparent[1]{}%
  }%
  \providecommand\rotatebox[2]{#2}%
  \newcommand*\fsize{\dimexpr\f@size pt\relax}%
  \newcommand*\lineheight[1]{\fontsize{\fsize}{#1\fsize}\selectfont}%
  \ifx\svgwidth\undefined%
    \setlength{\unitlength}{424.00089407bp}%
    \ifx\svgscale\undefined%
      \relax%
    \else%
      \setlength{\unitlength}{\unitlength * \real{\svgscale}}%
    \fi%
  \else%
    \setlength{\unitlength}{\svgwidth}%
  \fi%
  \global\let\svgwidth\undefined%
  \global\let\svgscale\undefined%
  \makeatother%
  \begin{picture}(1,0.76650943)%
    \lineheight{1}%
    \setlength\tabcolsep{0pt}%
    \put(0,0){\includegraphics[width=\unitlength,page=1]{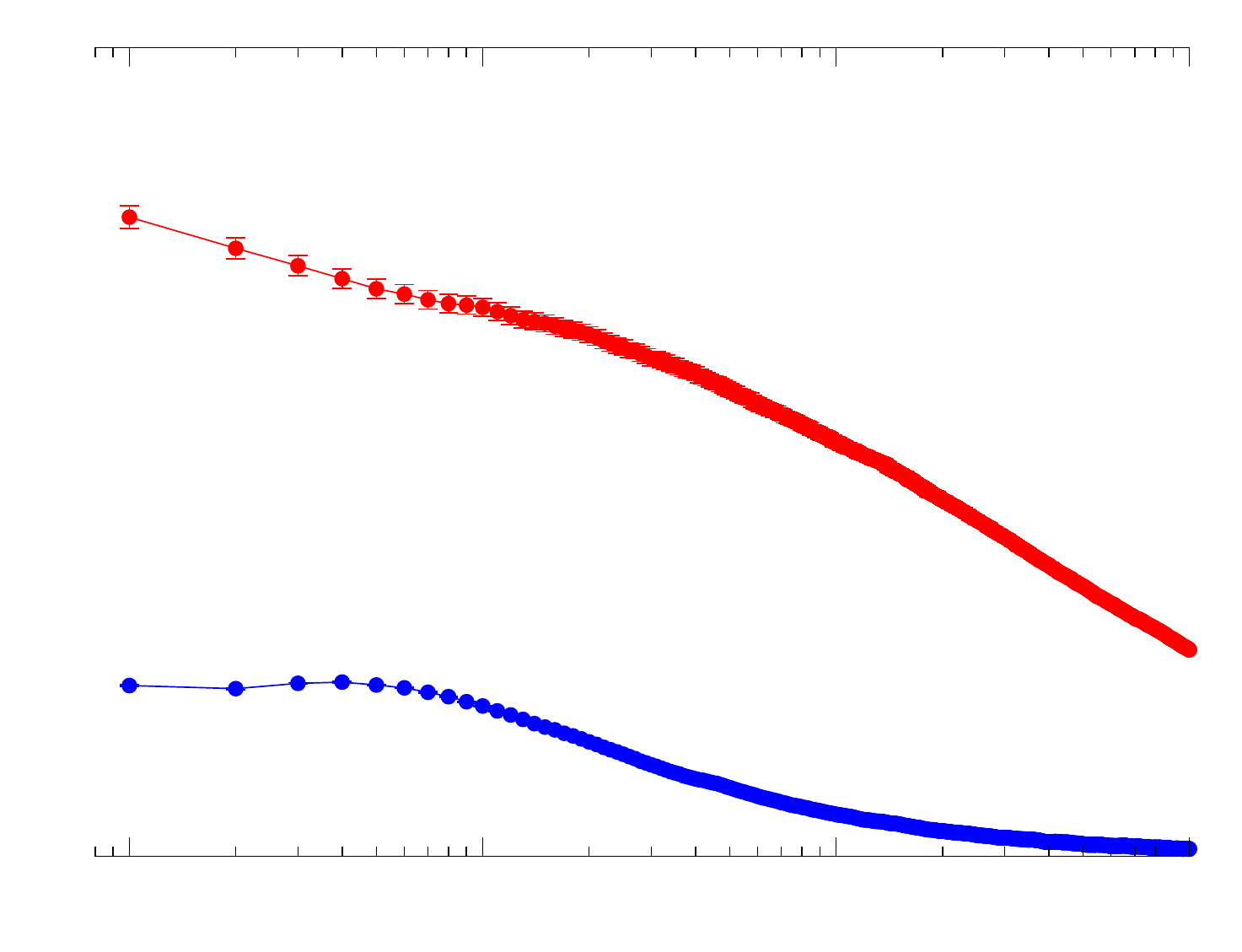}}%
    \put(0.0936217,0.03983491){\makebox(0,0)[lt]{\lineheight{1.25}\smash{\begin{tabular}[t]{l}\scriptsize $10^1$\end{tabular}}}}%
    \put(0.37353868,0.03983491){\makebox(0,0)[lt]{\lineheight{1.25}\smash{\begin{tabular}[t]{l}\scriptsize $10^2$\end{tabular}}}}%
    \put(0.65227642,0.03983491){\makebox(0,0)[lt]{\lineheight{1.25}\smash{\begin{tabular}[t]{l}\scriptsize $10^3$\end{tabular}}}}%
    \put(0.9321934,0.03983491){\makebox(0,0)[lt]{\lineheight{1.25}\smash{\begin{tabular}[t]{l}\scriptsize $10^4$\end{tabular}}}}%
    \put(0.53353868,0.01983491){\makebox(0,0)[lt]{\lineheight{1.25}\smash{\begin{tabular}[t]{l}$t$\end{tabular}}}}%
    \put(0,0){\includegraphics[width=\unitlength,page=2]{Kurtosis_2.pdf}}%
    \put(0.02804245,0.06957547){\makebox(0,0)[lt]{\lineheight{1.25}\smash{\begin{tabular}[t]{l}\scriptsize $0.0$\end{tabular}}}}%
    \put(0.02804245,0.61134033){\makebox(0,0)[lt]{\lineheight{1.25}\smash{\begin{tabular}[t]{l}\scriptsize $0.5$\end{tabular}}}}%
    \put(-0.01301887,0.390625){\rotatebox{0}{\makebox(0,0)[lt]{\lineheight{1.25}\smash{\begin{tabular}[t]{l}$\alpha_2(t)$\end{tabular}}}}}%
    \put(0,0){\includegraphics[width=\unitlength,page=3]{Kurtosis_2.pdf}}%
    \put(0.72825189,0.66037736){\makebox(0,0)[lt]{\lineheight{1.25}\smash{\begin{tabular}[t]{l}\tiny Percolation\end{tabular}}}}%
    \put(0,0){\includegraphics[width=\unitlength,page=4]{Kurtosis_2.pdf}}%
    \put(0.72825189,0.63089623){\makebox(0,0)[lt]{\lineheight{1.25}\smash{\begin{tabular}[t]{l}\tiny Solid Spheres\end{tabular}}}}%
    \put(0,0){\includegraphics[width=\unitlength,page=5]{Kurtosis_2.pdf}}%
  \end{picture}%
\endgroup%
\caption{Reduced kurtosis as a function of time for the percolation system and for the array of solid obstacles. The convergence to a Gaussian in 
supercritical percolation is very slow. 
\label{fig:Kurtosis_2}
} 
\end{figure}

To show the convergence to a Gaussian in three models of spacially disordered systems investigated (and to compare the speed of convergence for a potential landscape model and 
its mean-field counterpart, the CTRW) we consider the so-called Gaussian parameter (reduced kurtiosis \cite{Rahman_SM}),
\begin{equation}
 \alpha_2(t) = \frac{ \langle \mathbf{r}(t)^4  \rangle}{ a \langle \mathbf{r}(t)^2 \rangle^2 } - 1,
 \label{eq:nongauss}
\end{equation}
with a number constant $a$ which depends on the spacial dimension and is $a=2$ for two-dimensional cases discussed here, see e.g. \cite{Meroz2_SM}. 
The kurtosis is the integral measure of ``peakedness'' and ``fat-tailedness'' of the distribution.
For a Gaussian distribution the reduced kurtosis vanishes, and the departure of $\alpha_2(t)$ from zero gives a simple quantitative measure of ``non-Gaussianity''. 
Fig. \ref{fig:Kurtosis_1} represents the behavior of $\alpha_2$ as a function of time for the 
diffusivity landscape model and its mean field description given by a CTRW one. One readily infers that the PDF in the latter is not only closer to a Gaussian at any time (because it is effectively smoother), 
but also that the convergence at longer times takes place faster. 
Fig. \ref{fig:Kurtosis_2} shows the values of $\alpha_2(t)$ for the two other models, the diffusion on the infinite cluster in supercritical percolation, and the diffusion in 
the array of solid obstacles, again indicating the convergence to a Gaussian. Since the models are quite different, the two curves plotted are not intended for immediate comparison,
although one readily infers that the convergence to a Gaussian is fast for solid obstacles (presumably, due to fast convergence in the wings, as seen in Fig. 6 of the main text) and slow in percolation.

\subsection{h. A qualitative discussion of the percolation model}

At critical concentration (or more exactly, at $p_c + \epsilon$ with $\epsilon \to 0+$), the percolation cluster is a fractal object \cite{HavBen_SM}. The diffusion on this infinite cluster is anomalous, $\langle x^2 \rangle \propto t^\alpha$, with $\alpha = 2/d_w < 1$  with $d_w$ being the walk dimension. 
The central peak in the PDF does not seem to be reported before but is clearly visible in simulations in large clusters at $p_c$ (not shown).
The asymptotics of the PDF at criticality for $x$ large is essentially a stretched Gaussian one. 
The overall shape of the PDF at criticality corresponds on a logarithmic scale to a brace laying horizontally, like this $\overbrace{\;}$. 
The central ``chupchik'' may be a reminiscence of a non-analytical pre-exponential, as proposed for the diffusion on fractals, which e.g. persists for a Sierpinski gasket well into the asymptotic domain \cite{KZB_SM}. 

Above criticality, the convergence to a Gaussian is proven, and the estimates for the diffusion coefficient are known. 
It is generally accepted that at concentration slightly above percolation concentration the local structure of the infinite percolating cluster (at length scales $l \lesssim \xi$, with  $\xi$ being the correlation length of the percolation problem)
is similar to the one of the incipient percolation clusters at $p=p_c$, the same shape should apply to the intermediate asymptotics of the PDF (for $ x \ll \xi$) above criticality.

Note that in percolation, the survival probability for the  time first passage from the initial point to a point at distance $r$ scales $\Phi(t) \sim (t/\tau(r))^{1- \frac{d_s}{s}}$ 
\cite{Meroz_SM} with $d_s$ being the spectral dimension of the cluster, 
which is approximately $1.33$ and $\tau(r)$ is the corresponding characteristic time, scaling as $r^{d_w}$. In our case the distance $r$ has to be taken of the order of $\xi$.
The first passage to a boundary of a circular domain shows the same scaling due to the compact exploration property of random walks on the percolation cluster. Therefore the amount of the walks confined within the 
correlation length and not ``feeling'' that the system is not fractal at larger scales among all random walks decays approximately as $t^{-1/3}$, i.e. considerably slower than in the landscape model, as it is also clearly seen in simulations.  
Assuming that the central part of the PDF at criticality is mostly due to these trajectories, never leaving the initial patch, 
we conclude that its shape is retained by the central part of the distribution during the convergence to a Gaussian above the critical concentration, with stretched Gaussian tails converging to faster decaying Gaussian ones.

The qualitative explanation of the nature of the peak is very close to the one in the case of diffusivity landscapes. 
At criticality, the diffusion coefficient for normal diffusion $D \sim \langle x^2 \rangle /t$ on the infinite cluster vanishes in the long time 
limit, although the MSD is nonzero at any time. At any finite scale, this infinite cluster can be described as a diffusive medium where the 
local diffusion coefficient fluctuates strongly, has a mean close to zero, and a variance comparable with the mean. 
The authors were not able to find any previous information about the form of the distribution of these fluctuations, which would enable for a quantitative analysis, which is left for future work.


%